\documentclass[12pt]{article}
\usepackage{graphicx}
\usepackage{amsmath}
\def\Re{{\cal R \mskip-4mu \lower.1ex \hbox{\it e}\,}}
\def\Im{{\cal I \mskip-5mu \lower.1ex \hbox{\it m}\,}}
\def\ie{{\it i.e.}}
\def\eg{{\it e.g.}}

\def\etal{{\it et al.}}

\def\sub#1{_{\lower.25ex\hbox{$\scriptstyle#1$}}}
\def\tev{\,{\rm TeV}}
\def\gev{\,{\rm GeV}}

\def\to{\rightarrow}

\def\subw{_{\rm w}}
\def\mh{\ifmmode m\sbl H \else $m\sbl H$\fi}
\def\mch{\ifmmode m_{H^\pm} \else $m_{H^\pm}$\fi}
\def\mt{\ifmmode m_t\else $m_t$\fi}
\def\mc{\ifmmode m_c\else $m_c$\fi}
\def\mz{\ifmmode M_Z\else $M_Z$\fi}
\def\mw{\ifmmode M_W\else $M_W$\fi}
\def\mws{\ifmmode M_W^2 \else $M_W^2$\fi}
\def\mhs{\ifmmode m_H^2 \else $m_H^2$\fi}   
\def\mzs{\ifmmode M_Z^2 \else $M_Z^2$\fi}
\def\mts{\ifmmode m_t^2 \else $m_t^2$\fi}
\def\mcs{\ifmmode m_c^2 \else $m_c^2$\fi}
\def\mchs{\ifmmode m_{H^\pm}^2 \else $m_{H^\pm}^2$\fi}
\def\ztwo{\ifmmode Z_2\else $Z_2$\fi}
\def\zone{\ifmmode Z_1\else $Z_1$\fi}
\def\mtwo{\ifmmode M_2\else $M_2$\fi}
\def\mone{\ifmmode M_1\else $M_1$\fi}
\def\tb{\ifmmode \tan\beta \else $\tan\beta$\fi}
\def\xw{\ifmmode x\subw\else $x\subw$\fi}
\def\ch{\ifmmode H^\pm \else $H^\pm$\fi}
\def\lum{\ifmmode {\cal L}\else ${\cal L}$\fi}
\def\inpb{\ifmmode {\rm pb}^{-1}\else ${\rm pb}^{-1}$\fi}
\def\infb{\ifmmode {\rm fb}^{-1}\else ${\rm fb}^{-1}$\fi}
\def\epem{\ifmmode e^+e^-\else $e^+e^-$\fi}
\def\ppb{\ifmmode \bar pp\else $\bar pp$\fi}
\def\pbp{\ifmmode ~^(\bar p^)p\else $~^(\bar p^)p$\fi}
\def\bsg{\ifmmode B\to X_s\gamma\else $B\to X_s\gamma$\fi}
\def\bsll{\ifmmode B\to X_s\ell^+\ell^-\else $B\to X_s\ell^+\ell^-$\fi}
\def\bstt{\ifmmode B\to X_s\tau^+\tau^-\else $B\to X_s\tau^+\tau^-$\fi}

\newskip\zatskip \zatskip=0pt plus0pt minus0pt
\def\matth{\mathsurround=0pt}
\def\lsim{\mathrel{\mathpalette\atversim<}}
\def\gsim{\mathrel{\mathpalette\atversim>}}
\def\atversim#1#2{\lower0.7ex\vbox{\baselineskip\zatskip\lineskip\zatskip
  \lineskiplimit 0pt\ialign{$\matth#1\hfil##\hfil$\crcr#2\crcr\sim\crcr}}}

\renewcommand{\thefootnote}{\fnsymbol{footnote}}

\begin{document} \begin{titlepage} 
\rightline{\vbox{\halign{&#\hfil\cr
&SLAC-PUB-14454\cr
}}}
\vspace{1in} 
\begin{center}

{{\Large\bf Higgs Properties in the Fourth Generation MSSM:  Boosted Signals Over the 3G Plan}
\footnote{Work supported by the Department of 
Energy, Contract DE-AC02-76SF00515}\\}
\medskip
\medskip
\normalsize 
{\large R.C. Cotta, J.L. Hewett, A. Ismail, M.-P. Le and T.G. Rizzo{\footnote {e-mail: randoo, hewett, aismail, myphuong, rizzo@slac.stanford.edu}} \\
\vskip .6cm
SLAC National Accelerator Laboratory,  \\
2575 Sand Hill Rd, Menlo Park, CA 94025, USA\\}
\vskip .5cm

\end{center} 
\vskip 0.8cm

\begin{abstract} 

The generalization of the MSSM to the case of four chiral fermion generations (4GMSSM) can lead to significant changes in the phenomenology of the otherwise 
familiar Higgs sector. In most of the 3GMSSM parameter space, the lighter CP-even $h$ is $\sim 115-125$ GeV and mostly Standard Model-like while $H,A,H^\pm$ 
are all relatively heavy. Furthermore, the ratio of Higgs vevs, $\tan \beta$, is relatively unconstrained. In contrast to this, in the 4GMSSM, heavy fourth 
generation fermion loops drive the masses of $h,H,H^\pm$ to large values while the CP-odd boson, $A$, 
can remain relatively light and $\tan \beta$ is restricted to the range 
$1/2 \lsim \tan \beta \lsim 2$ due to perturbativity requirements on Yukawa couplings. We explore this scenario in some detail, concentrating on 
the collider signatures of the light  CP-odd Higgs at both the Tevatron and LHC. We find that while $gg \to A$ may lead to a potential signal in 
the $\tau^+\tau^-$ channel at the LHC, $A$  may first be observed in the $\gamma \gamma$ channel due to a highly loop-enhanced cross section that can be more 
than an order of magnitude greater than that of a SM Higgs for $A$ masses of $\sim 115-120$ and $\tan\beta<1$. We find that the CP-even states
$h,H$ are highly
mixed and can have atypical branching fractions.  Precision electroweak constraints, particularly 
for the light $A$ parameter space region, are examined in detail.

\end{abstract}

\renewcommand{\thefootnote}{\arabic{footnote}} \end{titlepage}


\section{Introduction and Background}
  
Although the Standard Model (SM) provides an excellent starting point from which to understand almost all experimental data, it provides
an incomplete picture of TeV scale physics as there are many questions it leaves unanswered. Four of the
most troubling of these questions
are ($i$) how is the hierarchy between the weak and Planck mass scales generated and stabilized, ($ii$) what is the nature of dark matter, ($iii$)
what generates the observed matter, anti-matter asymmetry, and ($iv$)
why are there three chiral fermion families? In order to address these issues, clearly some larger theoretical framework will be required.

Numerous theoretical scenarios have been suggested over the years to address these shortcomings of the SM, all of which have striking experimental 
signatures at the TeV scale\cite{Morrissey:2009tf}.
Supersymmetry (SUSY), in the guise of the Minimal Supersymmetric Standard Model (MSSM)\cite{Drees:2004jm}, provides one of the best motivated (and most popular)
frameworks in which to address both the hierarchy and dark matter problems and predicts a rich, testable
phenomenology.  The addition of a fourth family of chiral fermions remains attractive as a potential new source for the observed 
baryon asymmetry generated in the early universe\cite{Hou:2008xd} and as a way to  
address a number of potential issues in the heavy flavor sector\cite{Lunghi:2011xy}.  Although 
the MSSM with 3 chiral families of fermions (the 3GMSSM) has been relatively well explored, the four generation
MSSM has received relatively little attention except in the very recent literature\cite{Litsey:2009rp,Dawson:2010jx} where it has been found to have several interesting
features.  In particular, it has been noted\cite{Fok:2008yg} that the 4GMSSM with $\tan\beta$ near unity yields a strong first order phase transition.

In some ways, due to the totality of experimental constraints, the 4GMSSM parameter space is somewhat more restricted than 
the corresponding one of the 3GMSSM.
Only relatively recently has it been realized\cite{Kribs:2007nz} that a fourth chiral family of SM fermions remains allowed by the simultaneous requirements 
imposed by precision electroweak data\cite{lepew}, theoretical constraints on Yukawa coupling perturbativity\cite{Dawson:2010jx,DePree:2009ed}
and the direct search limits for the 
$\nu',l'$ leptons from LEP\cite{Achard:2001qw}
as well as the $b',t'$ quarks from both the Tevatron\cite{Lister:2011zt}
and now the LHC\cite{Chatrchyan:2011em}. Given these multiple constraints, 
the parameter space of allowed particle masses, particularly for the $b',t'$, is relatively restricted, and generally requires the $b',t'$ masses to 
lie in the 300-600 GeV range with mass splittings of order 50-100 GeV. A recent study of the 4GMSSM \cite{{Dawson:2010jx}}, shows that 
the experimental lower bounds on the $b',t'$ masses constrains 
the value of $\tan \beta$ such that it cannot differ very much from unity due to perturbativity requirements{\cite {CFH}}. Specifically $\tan\beta$ is
required to lie in the range $1/2 \lsim \tan \beta \lsim 2$.
One of the attractive features of the 4GMSSM is 
that the very large radiative corrections induced from loops involving the heavy fourth generation fermion masses allows one to push the lightest 
CP-even Higgs ($h$) mass far above the $\sim 130$ GeV conventional 3GMSSM upper bound, thus simultaneously relieving both fine-tuning issues as well 
as the direct Higgs search constraints. 

In this paper we will examine the properties of the 4GMSSM Higgs fields (such as mass spectrum, couplings 
and decay modes) and will begin to explore the collider physics of this Higgs sector. In particular we note the very interesting 
possibility that while large radiative corrections necessarily drive the CP-even ($h,H$) and charged Higgs ($H^\pm$) masses to large values 
$\gsim 350-400$ GeV, the CP-odd field ($A$) can remain relatively light with a mass in the 100-300 GeV range. Thus $A$ may be the lightest, and 
possibly, the first observable part of the Higgs sector of the 4GMSSM. Interestingly, such a light state easily avoids the usual LEP, Tevatron and 
LHC MSSM Higgs searches\cite{ewmoriond} since: ($i$) $A$, unlike $h$, does not couple to $WW^*$ or $ZZ^*$, so that searches for, \eg, $W+b\bar b$, $l^+l^-+$MET,
or $\gamma\gamma+$MET 
are trivially evaded, ($ii$) the sum of the $h$ and $A$ masses is forced to be rather large, $\gsim 400-500 $ GeV, so that associated production is 
absent or highly suppressed at colliders and ($iii$) since $\tan \beta$ is required to be close to unity in the 4GMSSM, constraints 
arising from searches for the $A\to \tau^+\tau^-$ final state are relatively easy to avoid. ($iv$) Furthermore, for low $\tan \beta$ and large $H^\pm$ 
masses, constraints from both $B\to \tau \nu$\cite{hfl} as well as top quark decays\cite{Aaltonen:2009ke}
are also easily satisfied. The state $A$ might, however, 
be observable in the $A\to \gamma \gamma$ decay mode at either the Tevatron or LHC if it is sufficiently light, especially as the values for 
both branching fractions $B(A\to gg,\gamma\gamma)$ can be significantly enhanced by the presence of the heavy fourth generation loop contributions.
In addition, we find that the $h$ and $H$ bosons are highly mixed states and become non-SM-like with atypical values for their branching fractions
into various final states. 

The paper is organized as follows.
In the next Section, we review the effects of the fourth generation on the radiative corrections for the MSSM Higgs sector and examine the resulting
Higgs mass spectrum. We also perform a global fit of the 4GMSSM to the precision electroweak data by analyzing the oblique electroweak parameters
$S,\, T$, and $U$ and determine the allowed range of parameter space for the special case of a light pseudoscalar Higgs.
We then study the collider phenomenology of the 4GMSSM Higgs sector, namely the Higgs production cross sections and branching ratios 
to various final states in Section 3. We compare these to present constraints from experiment and explore future detection prospects.  In particular,
we find that $gg\to A\to\gamma\gamma$ is a promising channel for early discovery.  Finally, we present our conclusions in Section 4.

\section{Radiative Corrections}

We begin our analysis by reviewing the effect of the radiative corrections to the Higgs sector arising from the fourth generation in the 4GMSSM. 
As noted by Ref.{\cite{Dawson:2010jx}}, since the fourth generation masses are so large, it suffices for our purposes to employ the one-loop, leading 
log effective potential approximation in performing these calculations\cite{Sher:1988mj}. In these computations, we must use as 
input the values of the $b',t',\nu'$ and $l'$ 
masses as well as the values of both $M_A$ and $\tan \beta$. In our analysis we take $\tan \beta$ to lie in the approximate range $1/2 \lsim \tan 
\beta \lsim 2$, as dictated by consistency with perturbative Yukawa couplings for fourth generation masses in the $\sim 300$ GeV range. 
In the limit where we neglect sfermion mixing and set all SUSY sfermion masses to a 
common value of $\sim 1$ TeV, only two further parameters must be specified: the common sfermion mass, $m_S$, and a common colorless gaugino mass, $m_\chi$.
Under these assumptions, we find that our conclusions are not much impacted by variations in these two parameters as our 
results are only logarithmically dependent on $m_S$, and the gaugino can potentially make only a rather small contribution 
to the rates for loop decays to the $\gamma \gamma$ final state for large masses. We note that the values of these input parameters must be chosen so as to satisfy 
all of the existing bounds from direct searches, precision electroweak data and the requirements of perturbative Yukawa couplings. 
The results presented below can, of course, be easily generalized to allow for both sfermion mixing as well as non-degenerate 
sparticle masses, but this will only modify the results we obtain in detail and not in any qualitative way.

To calculate the radiative corrections to the Higgs mass spectrum due to the addition of fourth generation fermions and their superpartners, we closely 
follow the work of Barger \etal\ in Ref. \cite{Barger:1991ed}. 
We stress that in performing these calculations both $M_A$ and $\tan \beta$ are to be treated as input parameters 
along with the masses 
of the fourth generation fermions and all the superpartners. In the general case, the masses associated with the CP-even Higgs fields are obtained
by diagonalizing the matrix

\begin{eqnarray}
M&=&\frac{1}{2}\left( \begin{array}{cc} \cot \beta & -1 \\-1 & \tan \beta \end{array} \right)M_{Z}^2\;\sin 2\beta \nonumber\\
& & +\frac{1}{2}\left( \begin{array}{cc}\tan \beta & -1 \\ -1 & \cot \beta\end{array} \right)M_A^2\;\sin 2\beta+\frac{g^2}{16\pi^2M_W^2}
\left( \begin{array}{cc}\Delta_{11} & \Delta_{12} \\ \Delta_{12} & \Delta_{22}\end{array} \right) \,,
\end{eqnarray}

\noindent where the $\Delta_{ij}$ are given by 
\begin{eqnarray}
\Delta_{11}&=&\sum_{(u,d)=(t',b'),(\nu',e')} \frac{N_c m_u^4}{\sin^2 \beta }~g_uC^2_u\mu^2 \nonumber\\
& & +\frac{N_c m_d^4}{\cos^2 \beta }\left[\ln\left(\frac{\tilde{m}^2_{d1}\tilde{m}^2_{d2}}{m_d^4}\right)+A_dC_d\left(2~\ln\left(\frac{\tilde{m}^2_{d1}}
{\tilde{m}^2_{d2}}\right)+A_dC_dg_d\right)\right]\,, \nonumber \\
\Delta_{22}&=&\Delta_{11} \left(\rm{with}\:\: u\leftrightarrow d\rm{,}\:\:\beta\rightarrow\frac{\pi}{2}-\beta\right)\,, \\
\Delta_{12}&=&\sum_{(u,d)=(t',b'),(\nu',e')} \frac{N_c m_u^4}{\sin^2 \beta }~\mu C_u\left[\ln\left(\frac{\tilde{m}^2_{u1}}{\tilde{m}^2_{u2}}\right)+
A_uC_ug_u\right]\nonumber \\
&  & + \left(u\leftrightarrow d\rm{,}\:\:\beta\rightarrow\frac{\pi}{2}-\beta\right) \,.\nonumber
 \end{eqnarray}

\noindent  $\tilde m_i$ are the physical sfermion masses, and the mixing parameters, $C_u$ and $C_d$, as well as the loop parameter $g_f$, are defined as

\begin{eqnarray}
C_{u}&\equiv&\frac{(A_{u}+\mu\;\rm{cot} \beta)}{(\tilde{m}^2_{u1}-\tilde{m}^2_{u2})}\,, \nonumber \\
C_{d}&\equiv&\frac{(A_{d}+\mu\;\rm{tan} \beta)}{(\tilde{m}^2_{d1}-\tilde{m}^2_{d2})}\,, \\
g_{f}&\equiv& 2-\frac{(\tilde{m}^2_{f1}+\tilde{m}^2_{f2})}{(\tilde{m}^2_{f1}-\tilde{m}^2_{f2})}\ln\left(\frac{\tilde{m}^2_{f1}}{\tilde{m}^2_{f2}}\right) \,.
\nonumber
\end{eqnarray}

In writing these expressions, we have assumed that there is no mixing between the fourth generation fermions or sfermions with their counterparts
in the other three 
generations. In particular, we specialize further to the case where the fourth generation squark/slepton mass eigenstates are the same as their interaction 
eigenstates, corresponding to $\mu=0$ and $A_{t',b',\nu',e'}=0$, wherein the mass matrix simplifies considerably. In our numerical results, we assume 
that all of the sfermions are degenerate with a mass of $m_S=1$ TeV. From these general expressions, we can obtain not only the contributions from 
the fourth generation, but also those from the usual top and bottom quarks. 

For the corresponding charged Higgs sector, we must diagonalize the analogous matrix 

\begin{eqnarray}
M&=&\frac{1}{2}\left( \begin{array}{cc} \tan\beta & 1 \\1 & \cot\beta\end{array} \right)M_{W}^2\;\sin 2\beta \nonumber\\
&+&\frac{1}{2}\left( \begin{array}{cc}\tan\beta & 1 \\ 1 & \cot\beta\end{array} \right)M_A^2\;\sin 2\beta+\frac{1}{2}\left( \begin{array}{cc}\tan
\beta & 1 \\ 1 & \cot\beta\end{array} \right)\widetilde{\Delta}\;\sin 2\beta \,,
\end{eqnarray}

\noindent where

\begin{eqnarray}
\widetilde{\Delta}&\equiv&\frac{g^2}{64\pi^2\;\sin^2\beta\;\cos^2\beta \;M_W^2} \nonumber\\
& & \times \sum_{(u,d)=(t',b'),(\nu',e')} ~N_c  ~\bigg(\frac{(m_d^2-M_W^2\;\cos^2\beta)(m_u^2-M_W^2\;\sin^2\beta)}{\tilde{m}^2_{u1}-\tilde{m}^2_{d1}}
\left[f(\tilde{m}^2_{u1})-f(\tilde{m}^2_{d1})\right] \nonumber\\
& & + \frac{m_u^2m_d^2}{\tilde{m}^2_{u2}-\tilde{m}^2_{d2}}\left[f(\tilde{m}^2_{u2})-f(\tilde{m}^2_{d2})\right]-\frac{2m_u^2m_d^2}{m_u^2-m_d^2}\left[f(m^2_u)-f(m^2_d)
\right]
	\bigg)\,,
\end{eqnarray}

\noindent with the function $f$ being given by $f(m^2)=2m^2\;\ln(m^2/M_W^2-1)$. 
Removing the Goldstone field $G^\pm$ leaves us with the desired mass (squared) of the charged 
Higgs field. As in the case of the neutral CP-even Higgs fields above, it is trivial to include the contributions from the ordinary third generation.

\begin{figure}[htbp]
\centerline{
\includegraphics[width=8.0cm,angle=90]{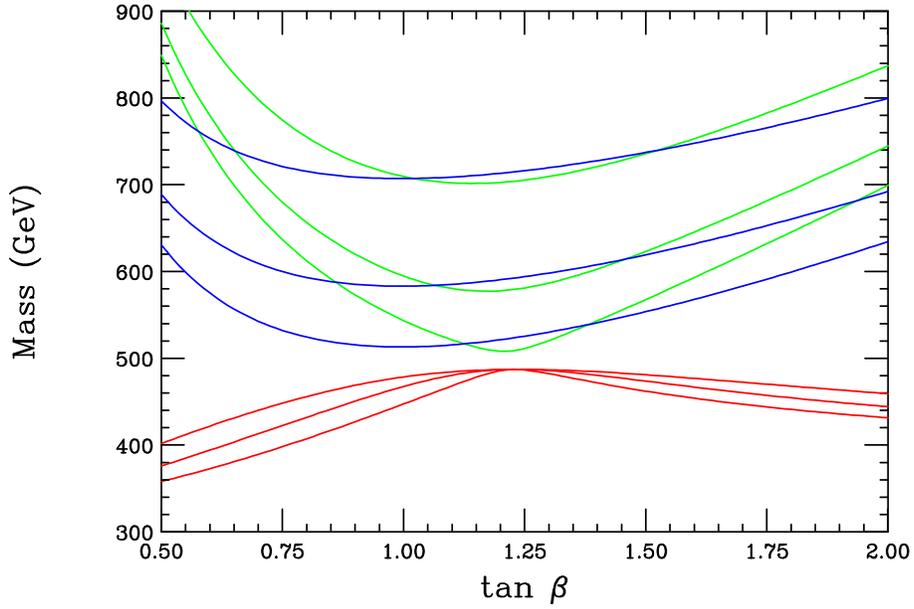}}
\vspace*{0.5cm}
\caption{Masses of the $h$(red), $H$(green) and $H^\pm$(blue) Higgs fields as functions of $\tan \beta$. The lower(middle, top) curve in each case 
corresponds to $M_A=115(300,500)$ GeV, respectively.  Here $m_{t'}=400$ GeV, $m_{b'}=350$ GeV, and $m_{l',\nu'}=300$ GeV with $m_S=1$ TeV have been assumed for 
purposes of demonstration.}
\label{masses}
\end{figure}

The primary results of this analysis are the masses of the $h,H$ and $H^\pm$ fields as functions of the input parameters. Figure~\ref{masses} shows a 
representative sample mass spectrum for these particles as a function of $\tan \beta$ for three different values of $M_A$ (115, 300, and 500 GeV) 
and taking $m_{t'}=400$ GeV,
$m_{b'}=350$ GeV and $m_{\ell',\nu'}=300$ GeV.  As can be seen in the formulae given above, the CP-even Higgs masses are expected to grow approximately
quadratically with the fourth generation mass scale (with the other parameters being held fixed).  This expectation was verified explicitly 
in {\cite {Dawson:2010jx}} where the sensitivity to variations in the fourth generation fermion masses was examined and we obtain similar results here. 
In this Figure, we observe 
that ($i$) the mass of $h$ is not particularly sensitive to the value of either $M_A$ or $\tan \beta$ and is primarily driven only by the masses of the fourth 
generation particles.  
($ii$) The values of $M_{H}$ are found to be sensitive to both of the input parameters. 
($iii$) On the other hand, $M_H^\pm$, while not particularly sensitive to the value of $\tan \beta$, does vary with $M_A$.
For these choices of fourth generation masses we see that the CP-even states $h$ and $H$ are quite heavy and thus it is easy for 
$A$ to be the lightest member of the Higgs spectrum and so it, perhaps, might be most easily observed at the Tevatron or LHC. Note that in all cases
the $H^\pm$ boson is too heavy to play much of a role in flavor physics, particularly since $\tan \beta$ is always near unity.

As the 4GMSSM includes many new electroweak states beyond those of the SM, one must carefully consider the effect that these states will have on the 
precise measurements of the electroweak interactions that are seen to be consistent with the SM (with a light SM Higgs, $m_h\sim100\gev$). 
4GMSSM scenarios with a light $A$ boson (\ie, $M_A< 300$ GeV) and/or $\tan \beta <1$ have not been previously considered so it behooves us to re-examine these 
cases. Here we focus on oblique corrections to the S, T and U parameters\cite{Peskin:1991sw} from the 4GMSSM with $M_A=115 \gev$ and 
$0.6<\mathrm{tan}\beta<1.8$; a broader and more detailed investigation of such corrections in the context of the 4GMSSM has been presented in 
\cite{Dawson:2010jx}. 
	
We compute the fourth generation fermion and Higgs sector contributions to the $S\,, T$ and $U$ parameters following the formulae in \cite{He:2001tp}. We 
neglect sfermion contributions as we assume all sfermions are heavy and degenerate, having $M_{\rm{SUSY}}\sim1\tev$, and hence their contributions are negligible. 
Fermion and Higgs contributions to the $U$ parameter, while non-zero, are also negligibly small in the parameter space considered here. The contributions 
due to the fermions alone were found to be numerically consistent with the results \cite{Kribs:2007nz}.
	
Constraints on new corrections to the $S\,, T$ and $U$ parameters are experimentally determined to be \cite{pdg}
\begin{eqnarray}
\Delta S=S-S_{SM}&=&-0.08\pm0.10 \nonumber \\
\Delta T=T-T_{SM}&=&0.09\pm0.11 \nonumber \\
\Delta U=U-U_{SM}&=&0.01\pm0.10,
\label{deltaSTU}
\end{eqnarray}
	
\noindent where the values above correspond to subtracting SM contributions which are calculated at the reference scale{\footnote{We note that while
$m_h$ can vary between approximately 360-500 GeV as the 4GMSSM parameter space is varied, we observe that the use of data values centered around the
reference point $m_{h,ref}$=300 GeV does not lead to 
any significant shift in the allowed regions displayed in the figure below.}}  $m_{h,ref}=300\gev$. 
The corrections $\Delta S$, $\Delta T$ and $\Delta U$ come purely from new physics, $\ie$, the SM contributions (with $m_h=300\gev$) to $\Delta S$, 
$\Delta T$ and $\Delta U$ are zero, in reasonable agreement with the above experimental ranges. We determine a $\Delta\chi^2$ value for points in 
4GMSSM space, following \cite{Dawson:2010jx},

\begin{equation}
\Delta\chi^2=\sum_{ij}(\Delta X_i-\Delta\widehat{X}_i)(\sigma_{ij})^{-1}(\Delta X_j-\Delta\widehat{X}_j),
\end{equation}

\noindent where the $\Delta\widehat{X}_i$ are the central values $\Delta S$, $\Delta T$ and $\Delta U$ of Eqn.\ (\ref{deltaSTU}), the $\Delta X_i$ are the 
fourth generation fermion and Higgs contributions to $\Delta S$, $\Delta T$ and $\Delta U$ from the particular 4GMSSM model and $\sigma_{ij}=
\sigma_i\rho_{ij}\sigma_{j}$ is the covariance matrix built from the errors $\sigma_i$ in Eqn.\ (\ref{deltaSTU}) and from

\begin{equation}
\rho=\left( \begin{array}{ccc} 1.0 & 0.879 & -0.469 \\ 0.879 & 1.0 & -0.716 \\ -0.469 & -0.716 & 1.0\end{array} \right) \,.
\end{equation}

\noindent Fits with $\Delta\chi^2>7.815$ would then be excluded at 95\% confidence by precision electroweak measurements.

\begin{figure}[htbp]
\centerline{
\includegraphics[width=16.0cm]{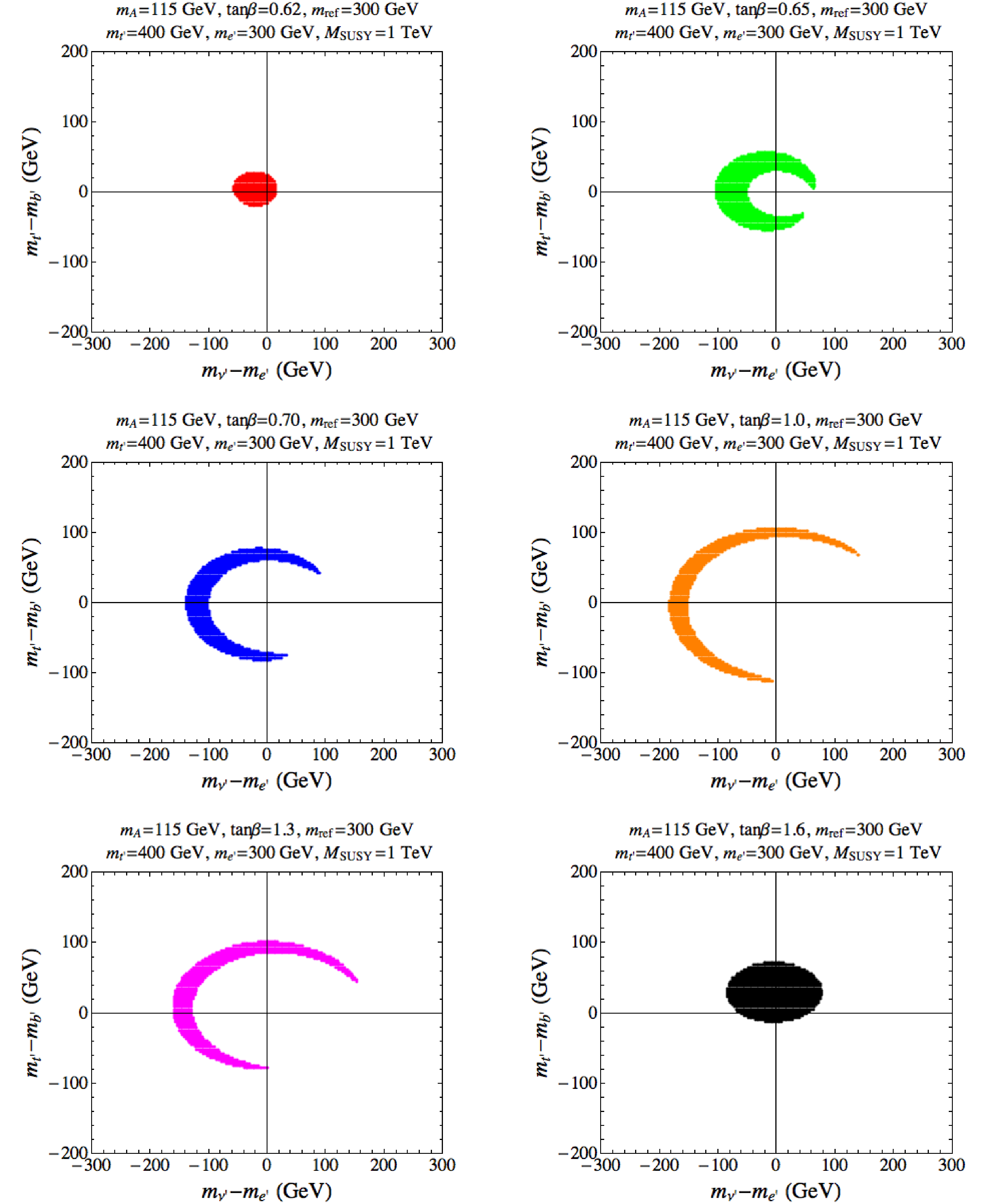}}
\vspace*{0.5cm}
\caption{Points representing particular 4GMSSM models allowed by precision electroweak constraints ($\Delta\chi^2<7.815$) are displayed in 
the $(m_{t'}-m_{b'})$ vs.\ $(m_{\nu'}-m_{e'})$ plane. In all cases we take $m_A=115\gev$, $M_{\rm{SUSY}}\sim1\tev$, $m_{t'}=400\gev$ and $m_{e'}=300\gev$. 
Points in distinct panels correspond to models with distinct values of $\mathrm{tan}\beta$: $\mathrm{tan}\beta=0.62$ (red), $\mathrm{tan}\beta=
0.65$ (green), $\mathrm{tan}\beta=0.70$ (blue), $\mathrm{tan}\beta=1.0$ (orange), $\mathrm{tan}\beta=1.3$ (magenta), $\mathrm{tan}\beta=1.6$ (black).}
\label{figs:stu}
\end{figure}

In Figure \ref{figs:stu} we display points in the $(m_{t'}-m_{b'})$ vs.\ $(m_{\nu'}-m_{e'})$ plane that are allowed by precision electroweak 
measurements  and consistent with unitarity (for $m_{t'}=400\gev$ and $m_{e'}=300\gev$, this means $m_{b'}<525\gev$ and 
$m_{\nu'}<750\gev$ \cite{Dawson:2010jx}). We observe that for  $\mathrm{tan}\beta\sim 1$, there is a relatively tuned set of fourth generation 
doublet splittings that are consistent with precision constraints, while, for somewhat larger and smaller values of $\mathrm{tan}\beta$, 
small splittings (or even degenerate doublets) are required for the 4GMSSM to be consistent with the precision electroweak data.


%

Note that since the $t'$ and $\ell'$ masses as well as $M_A$ are being held constant in these figures, the variation with $\tan \beta$ arises from only two 
unique sources: the changes in the Higgs couplings to the fermions and gauge bosons described above and the corresponding changes in the various Higgs boson 
mass splittings entering the loop functions. Since the mass splitting between the Higgs fields is greatest at the two ends of the allowed $\tan \beta$ range, 
we see that in such cases the allowed region in the fourth generation mass splitting plane then reduces to a solid ellipse. Furthermore, when these mass 
splittings are minimized for $\tan \beta \simeq 1-1.2$ we see that the allowed arc-shaped region in this plane has its maximal radial extent.

\section{Collider Phenomenology}

We next examine the collider phenomenology of the 4GMSSM Higgs sector, paying particular attention to the region of parameter space that results in different
signatures from the three generation case.  Throughout this section we shall assume $m_{t'}=400$ GeV,
$m_{b'}=350$ GeV, $m_{\ell',\nu'}=300$ GeV and $m_S=1$ TeV in presenting our results.  We find that varying the fourth generation fermion masses within
their allowed ranges does not qualitatively modify our conclusions.

\begin{figure}[htbp]
\centerline{
\includegraphics[width=6.0cm,angle=90]{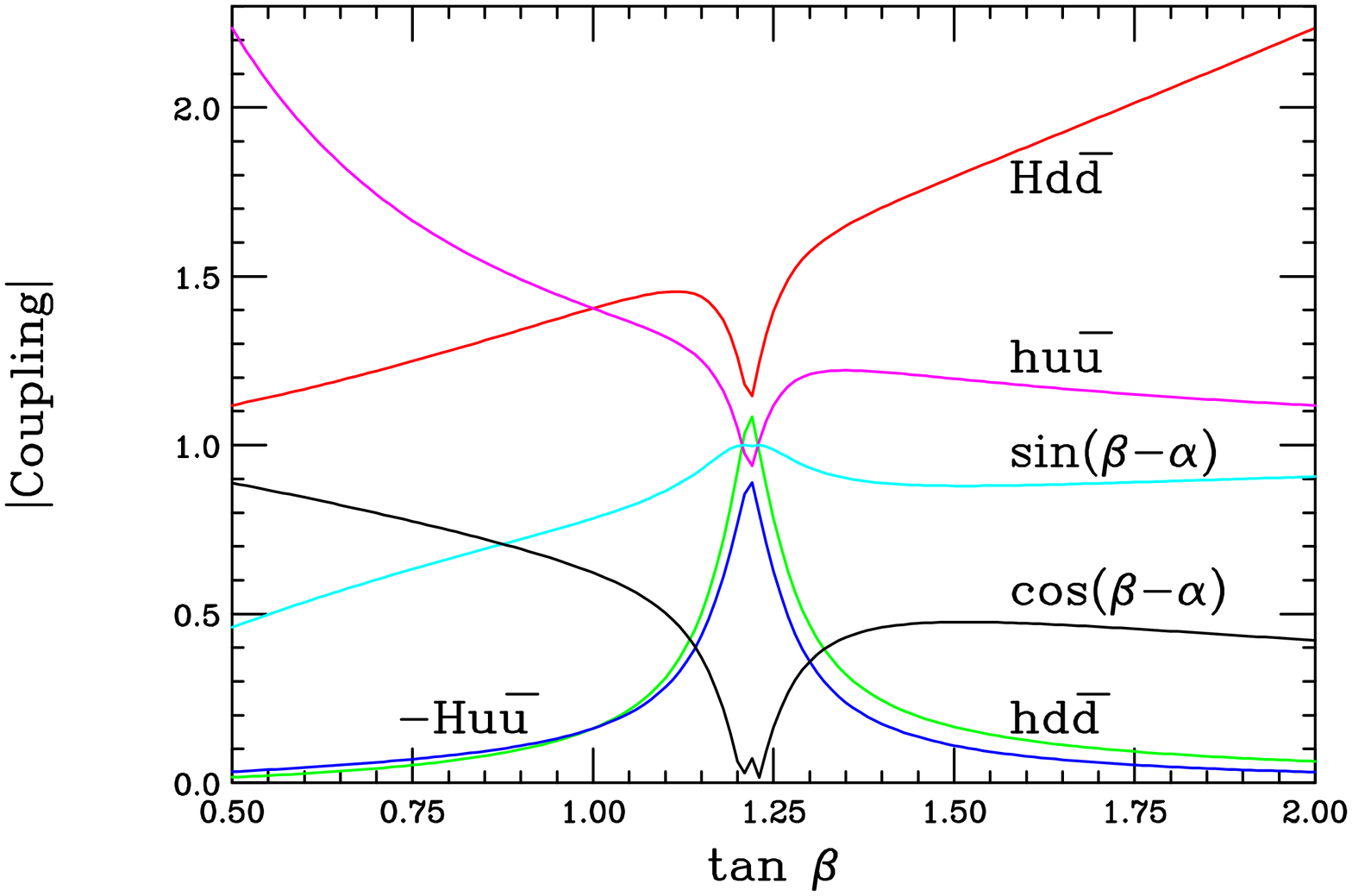}
\hspace*{5mm}
\includegraphics[width=6.0cm,angle=90]{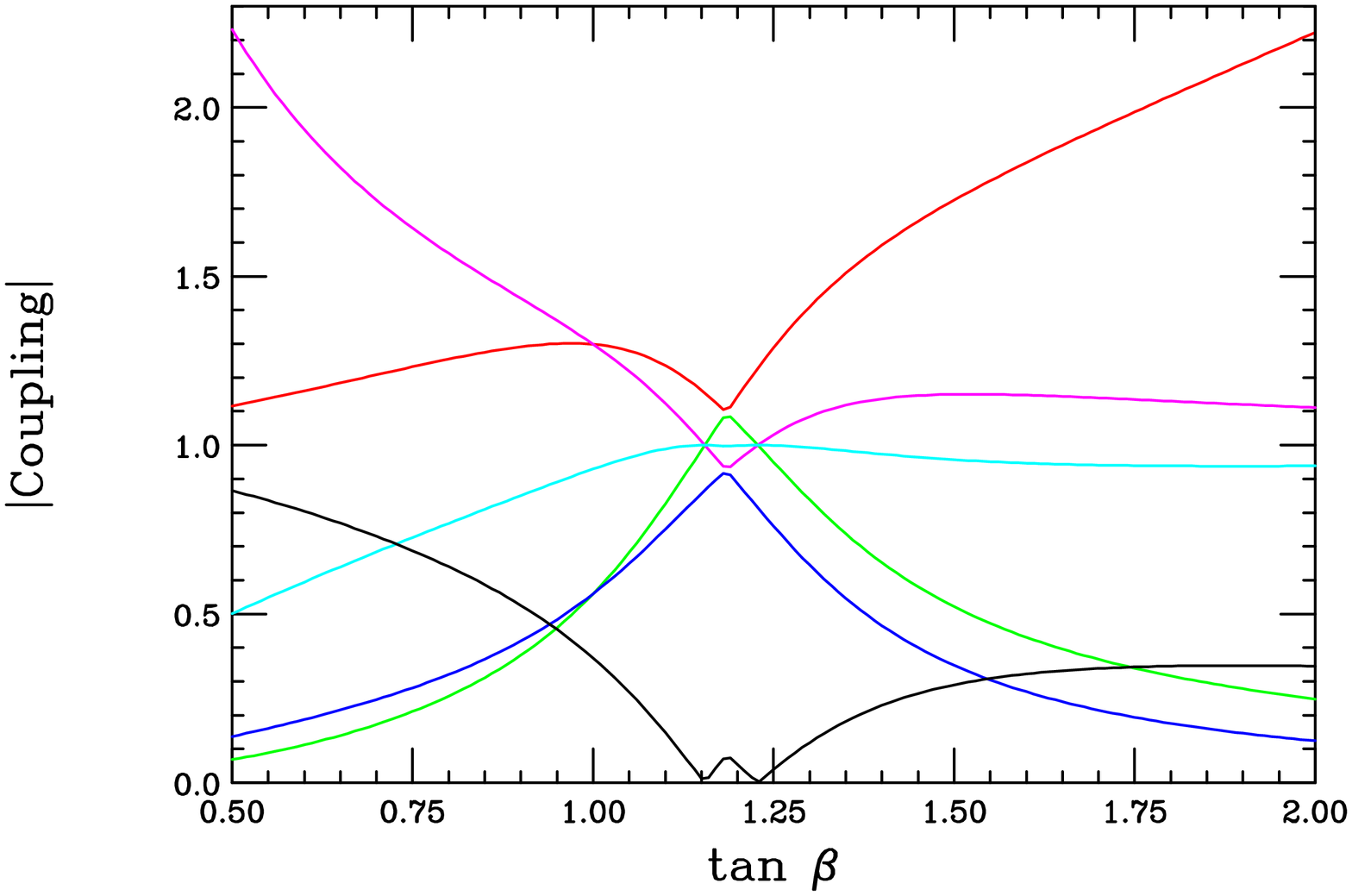}}
\vspace*{0.5cm}
\centerline{
\includegraphics[width=6.0cm,angle=90]{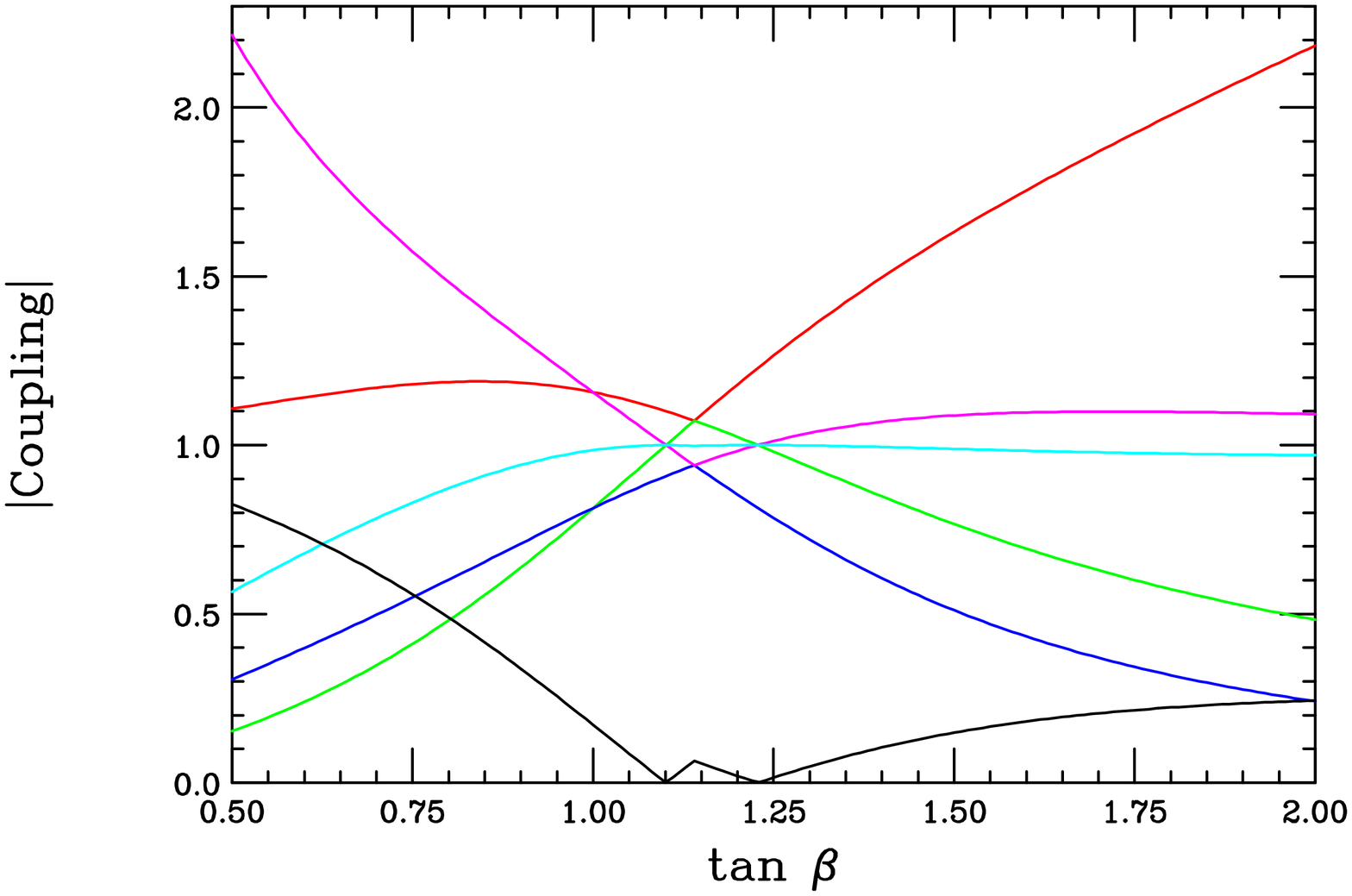}}
\vspace*{0.5cm}
\caption{CP-even Higgs boson coupling factors, normalized to the corresponding SM Higgs couplings, as a function of $\tan \beta$ for $M_A=115(300,500)$ in the top 
left(top right, bottom) panel. Here $m_{t'}=400$ GeV, $m_{b'}=350$ GeV, and $m_{l',\nu'}=300$ GeV with $m_S=1$ TeV have been assumed for purposes 
of demonstration. All curves are as labeled in the upper left-hand panel.}
\label{couplings}
\end{figure}

Our first step is to determine the various coupling coefficients for the $h,H$ bosons to the $u,d$-type fermions 
and SM gauge bosons as functions of $\tan \beta$ and $M_A$ for our fixed values of the other input parameters. The corresponding couplings of the
pseudoscalar $A$ boson to the  
fermions are simply given by $\tan \beta$ and its inverse, and $VVA$-type couplings are absent. The form of these couplings follow directly from the equations 
describing the radiative corrections to the Higgs sector in the previous section
with the diagonalization of the CP-even Higgs mass matrix then determining the mixing angle $\alpha$. Figure.~\ref{couplings} shows these various 
couplings as functions of $\tan \beta$ for three different values of $M_A$. These couplings display a strong
$\tan \beta$ dependence in the range of interest, while showing 
only a somewhat mild dependence on $M_A$ except for an overall broadening of the peak observed in the center of the figures near $\tan \beta \sim 1.2$ as
the value of $M_A$ is increased. 
Interestingly, we find that for a substantial fraction of the range of $\tan \beta$, the CP-even Higgs fields have significant mixing so that neither $h$ nor 
$H$ are SM-like. This is in contrast to the usual scenario in the 3GMSSM. 
Note that generally $h(H)$ has stronger(weaker) couplings to $u\bar u$-type quarks than does the SM Higgs while the reverse is found to 
be true for the corresponding $d\bar d$-type couplings. Also note that it is possible for both $h$ and $H$ to simultaneously have substantially large couplings 
to the SM $W,Z$ bosons.

\begin{figure}[htbp]
\centerline{
\includegraphics[width=6.0cm,angle=90]{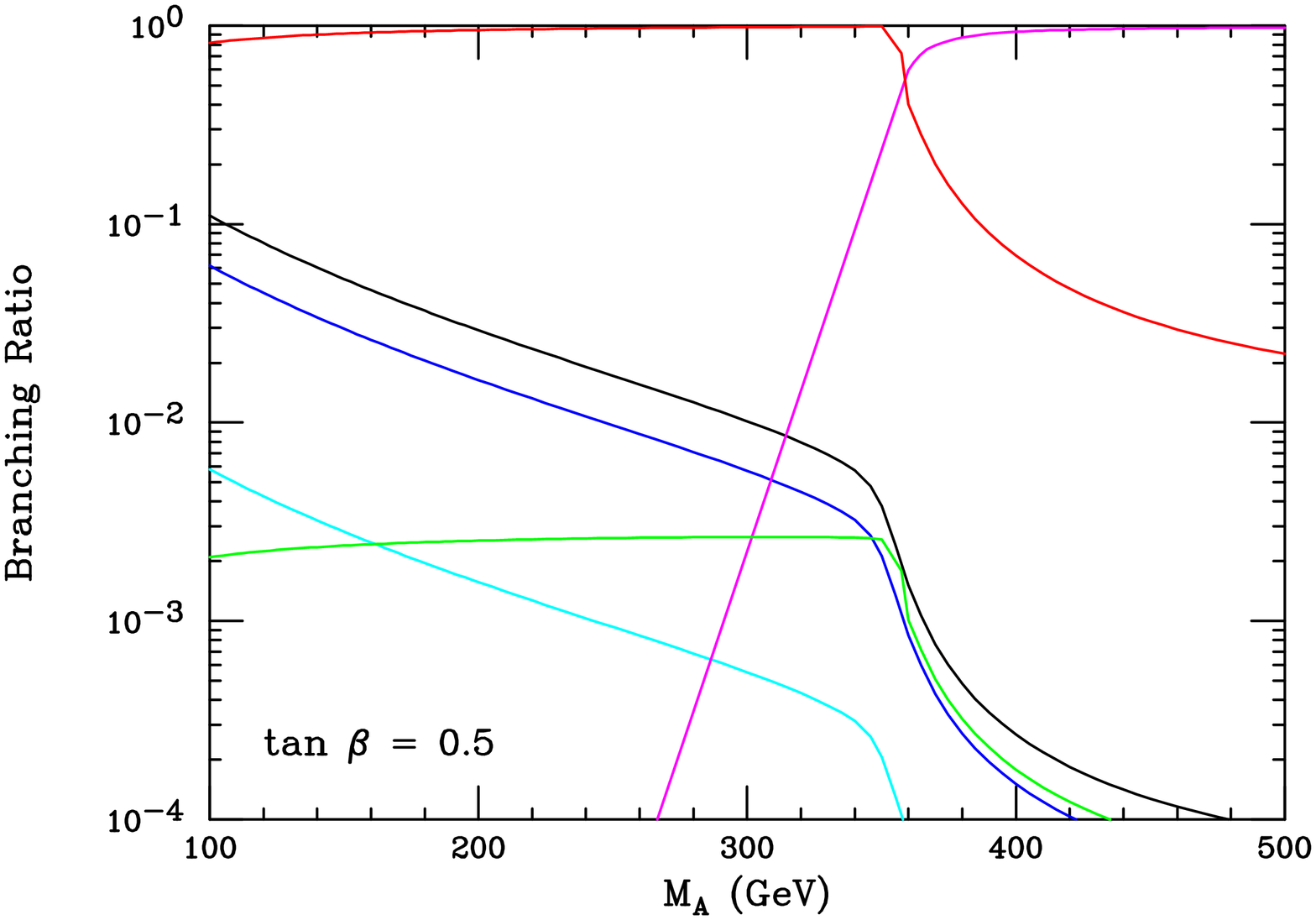}
\hspace*{5mm}
\includegraphics[width=6.0cm,angle=90]{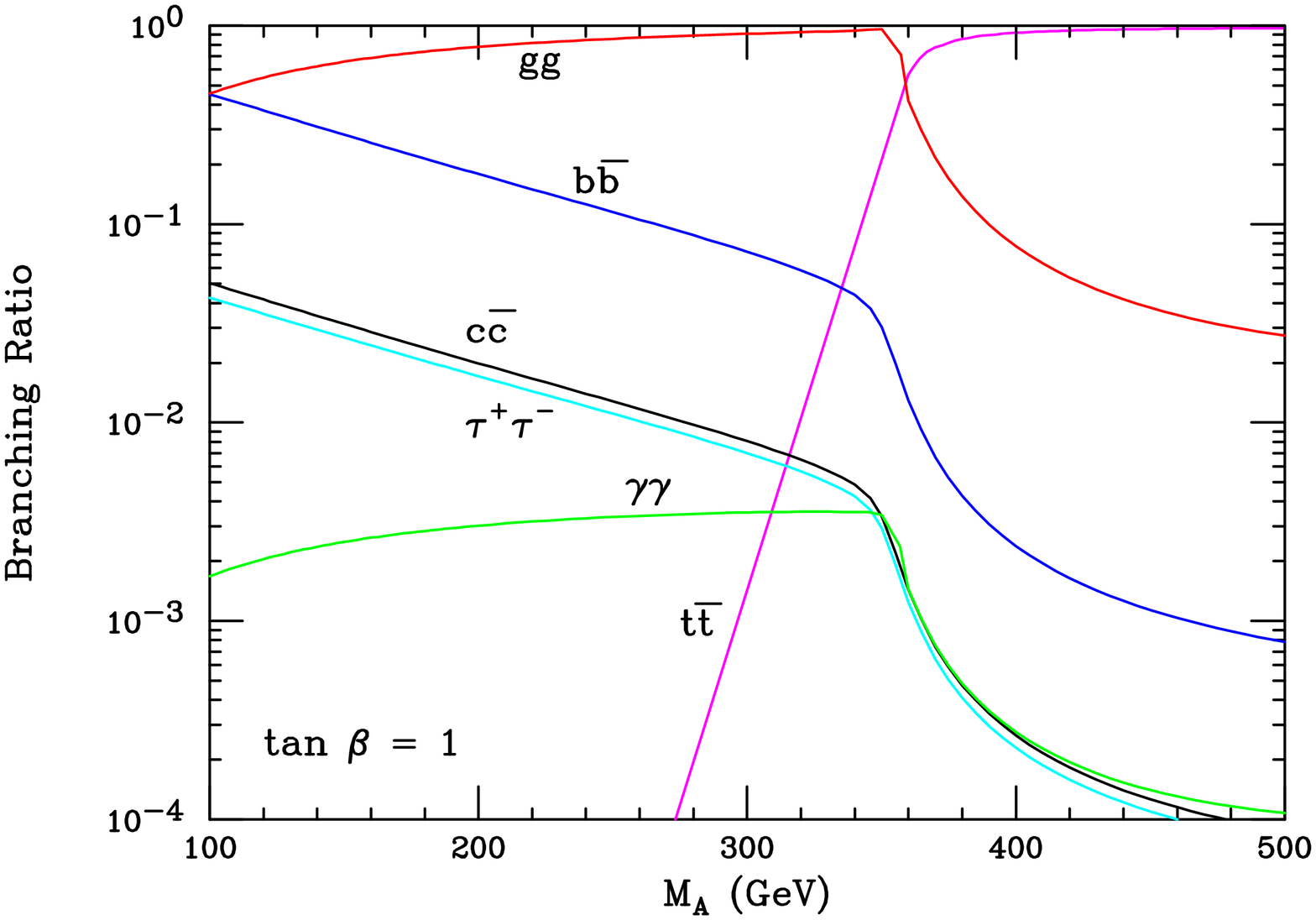}}
\vspace*{0.5cm}
\centerline{
\includegraphics[width=6.0cm,angle=90]{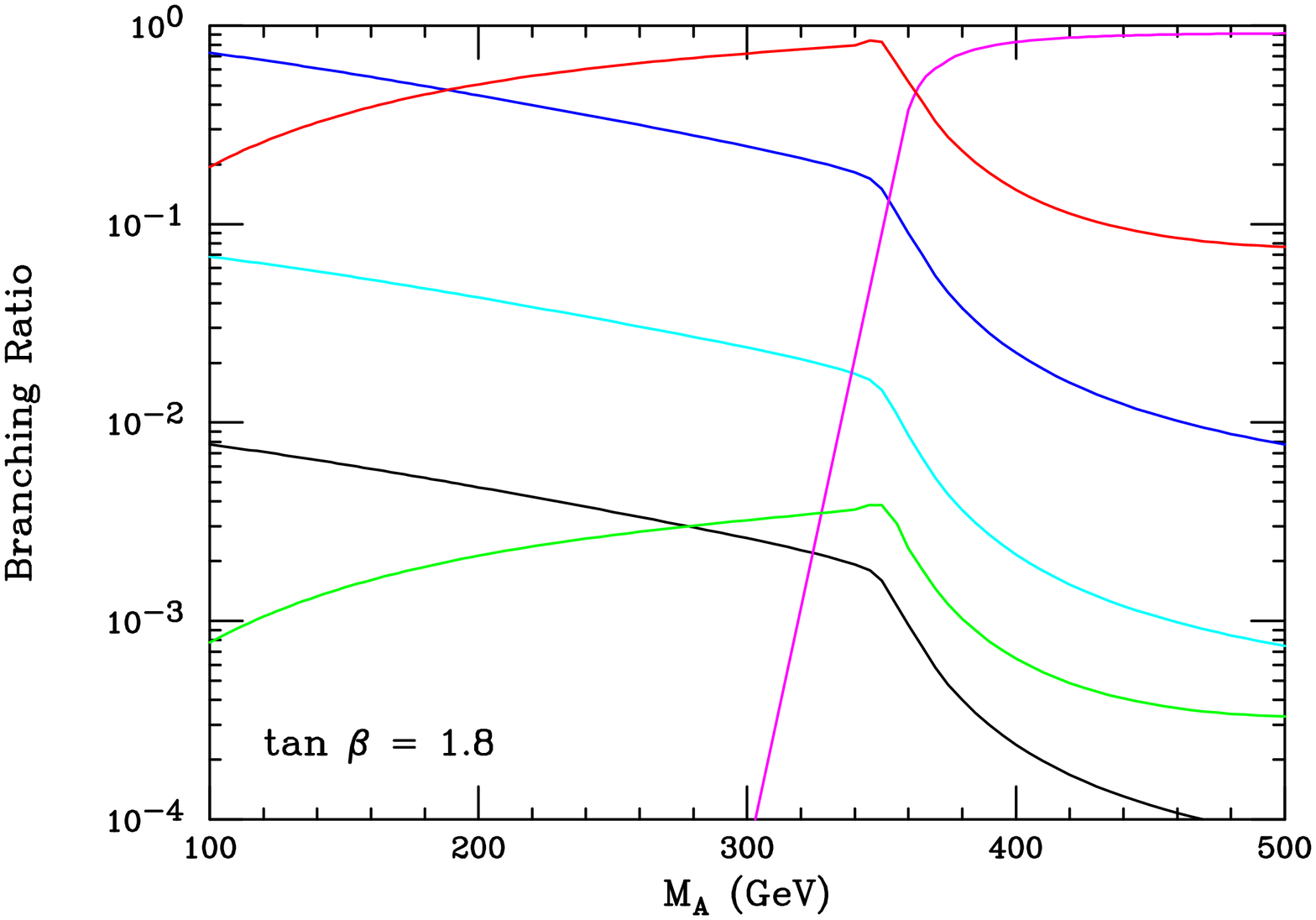}}
\vspace*{0.5cm}
\caption{Branching fractions of the CP-odd state $A$ as a function of $M_A$ for the same input masses as in the previous figure. The top left(right) 
panel assumes $\tan \beta =0.5(1)$ while the bottom panel assumes $\tan \beta =1.8$. All curves are as labeled in the upper right-hand panel.}
\label{abfs}
\end{figure}

Once the couplings of the various Higgs states are determined, we can calculate their respective branching fractions. Here, we first pay special attention to 
the CP-odd field $A$ since it may be the lightest of the Higgs states. Figure~\ref{abfs} shows these branching fractions as a function of $M_A$ for three different 
values of $\tan \beta$ taking the fourth generation masses as above. Note that the channel $A\to gg$ is greatly enhanced for $M_A<2m_t$, and is the dominant
decay mode for $\tan\beta<1$. The $\gamma\gamma$ partial width is also found to be enhanced by up to a factor of two over that of the SM Higgs, but this
increase is found to wash out in the branching fraction. Together, this can lead to large signal rates for $gg\to A \to 
\gamma \gamma$ as will be discussed below. Note that the size of the $b,c,\tau$ branching fractions are particularly sensitive to the value of $\tan \beta$. 
For larger values of $M_A$, the $\tau^+\tau^-$ and $\gamma\gamma$ channels are roughly comparable. 

\begin{figure}[htbp]
\centerline{
\includegraphics[width=6.0cm,angle=90]{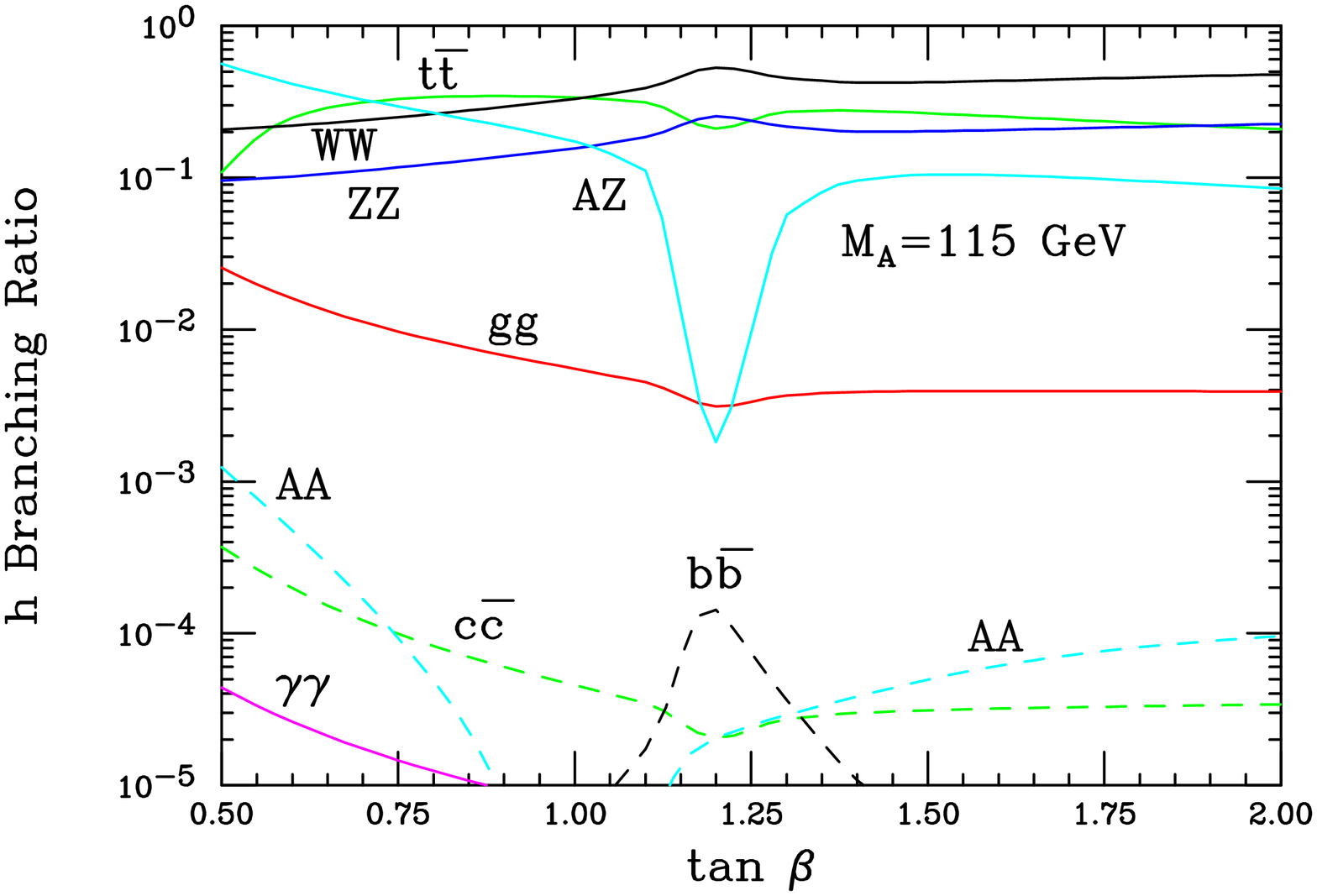}
\hspace*{5mm}
\includegraphics[width=6.0cm,angle=90]{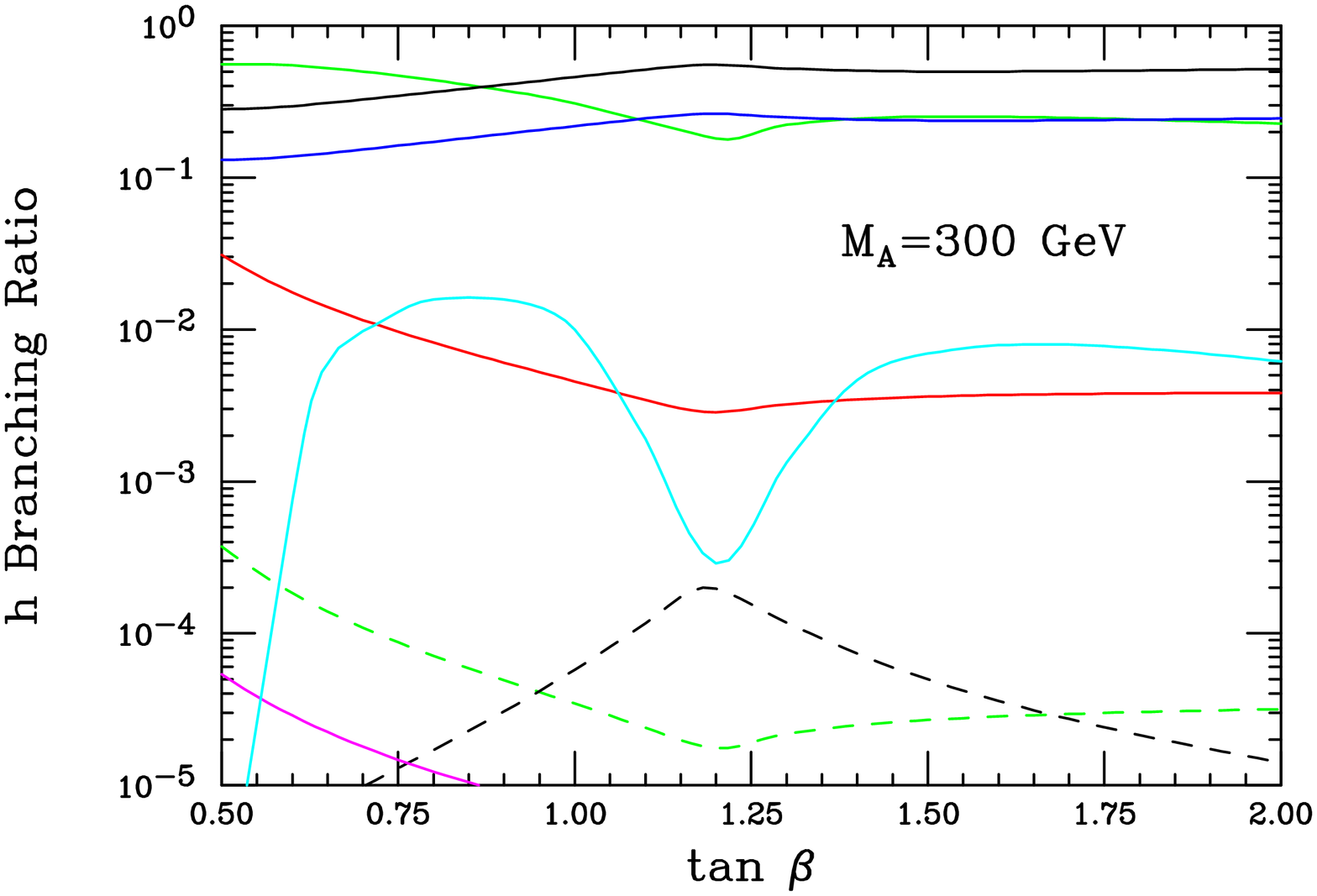}}
\vspace*{0.5cm}
\centerline{
\includegraphics[width=6.0cm,angle=90]{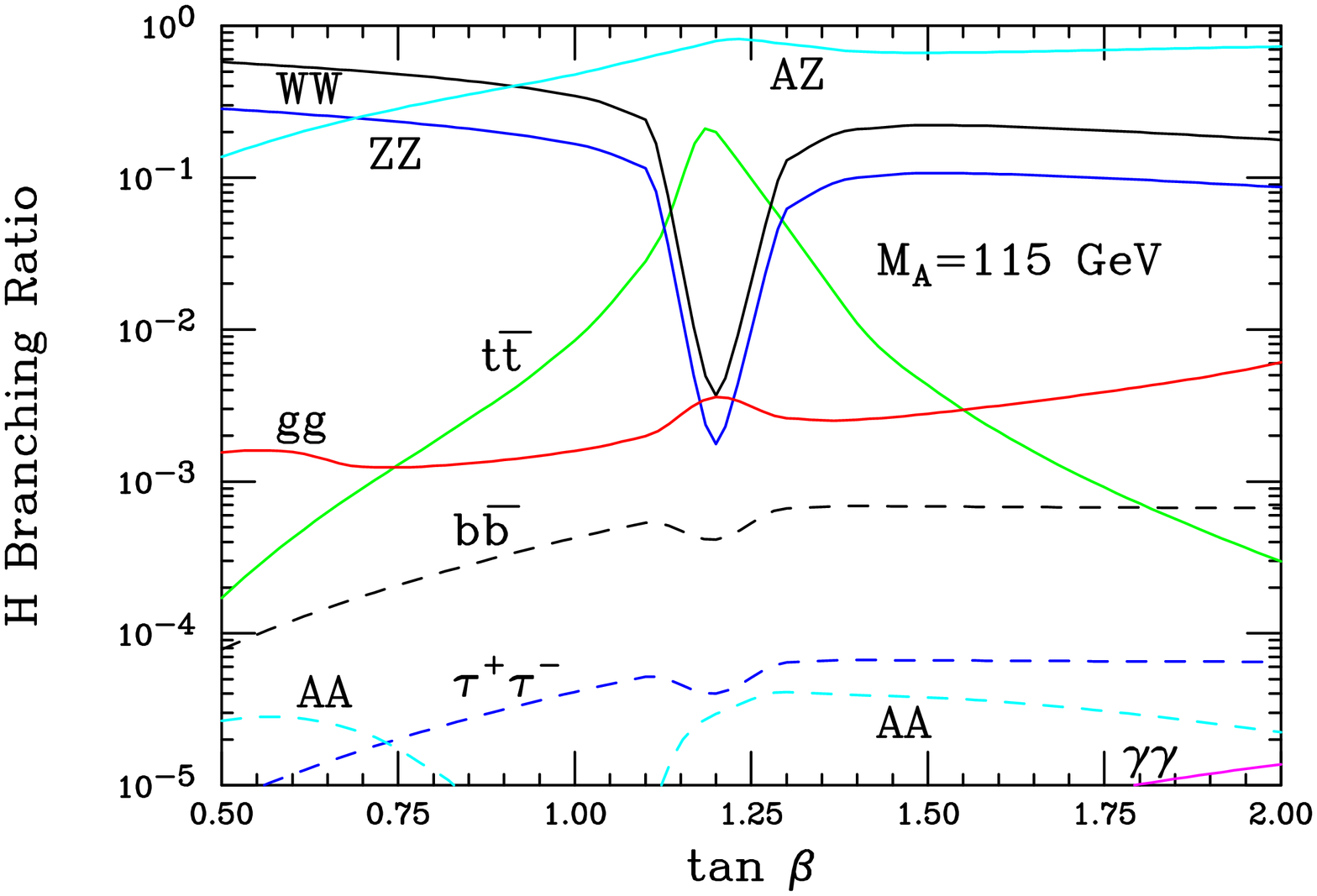}
\hspace*{5mm}
\includegraphics[width=6.0cm,angle=90]{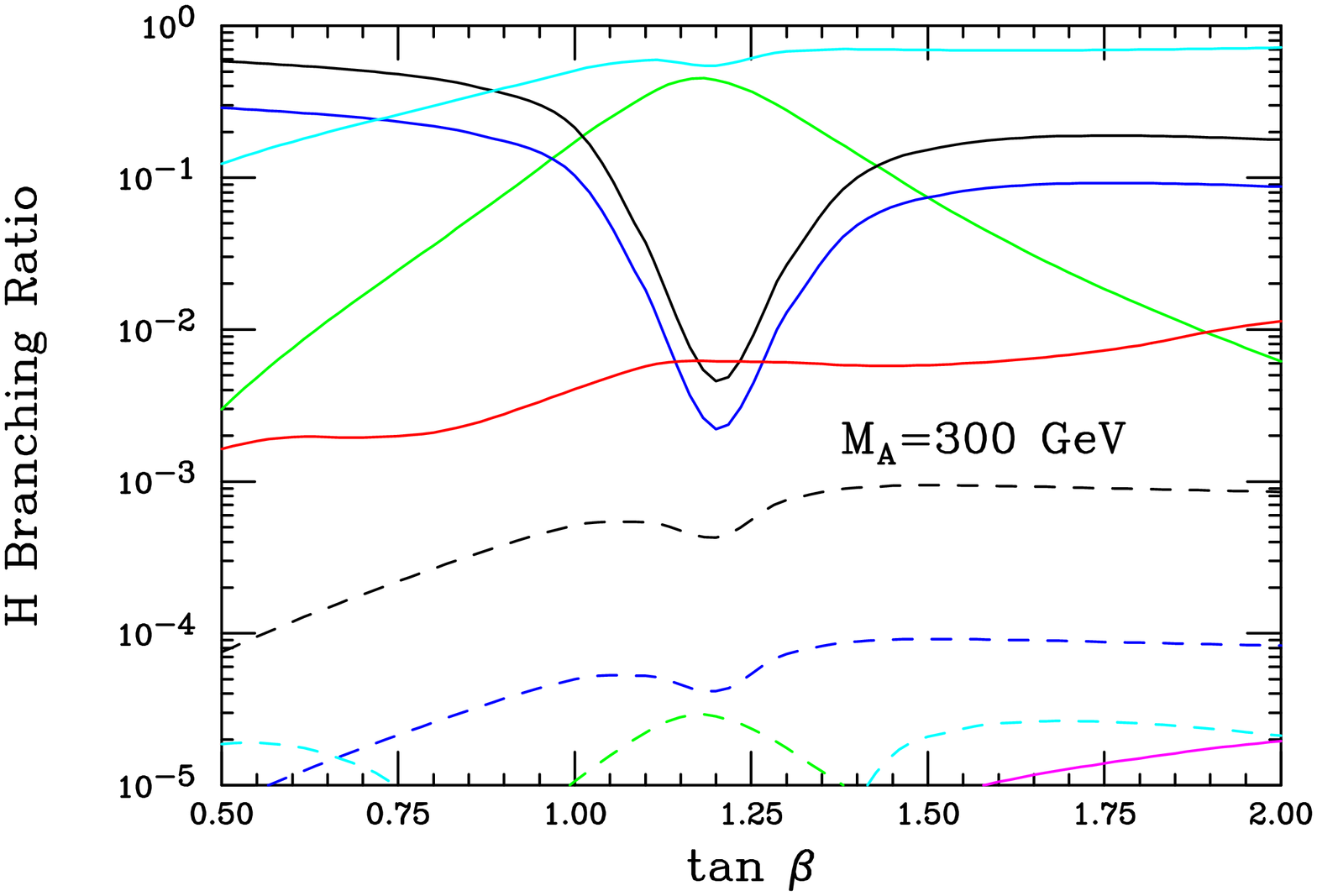}}
\vspace*{0.5cm}
\caption{Branching fractions for the $h$(top) and $H$(bottom) as functions of $\tan \beta$ for $M_A=115(300)$ GeV in the left(right) panels. The 
other input masses are taken to be those as employed above. The curves in the right panels correspond to the same decays as the ones in the left panels.}
\label{hHbfs}
\end{figure}

Turning to the CP-even Higgs bosons, Fig.~\ref{hHbfs} shows the relevant branching fractions.  Here, we have assumed for simplicity 
that decays to pairs of fourth generation fermions are not kinematically allowed.{\footnote{Decays
to fourth generation fermions are not kinematically allowed for the lightest Higgs boson $h$.  However,
for fixed $M_A$, as $\tan\beta$ is varied, decay channels to the fourth generation may open up for the heavier $H$ boson if the 4G fermion masses
are light enough.}}  As expected, $h$ and $H$ decays to $VV$ (with $V$ being either the SM $W$ or $Z$ boson) can 
dominate over most of the parameter space.  In the case of $h$, the $t\bar t$ mode 
is of comparable importance. For $\tan \beta \sim 1.2$, as can be seen from Fig.~\ref{couplings}, $h$ becomes more SM-like and, hence, $H$ 
nearly decouples from $VV$ in this region. This is reflected in the dip in the $H\to VV$ branching fractions near this particular $\tan \beta$ value. Similarly, 
since the $Hu\bar u$ coupling is usually suppressed relative to the corresponding SM value (except again near $\tan \beta \sim 1.2$), the $H\to t\bar t$ 
decay is generally found to be sub-dominant. $h,H$ branching fractions to both $b\bar b$ and $\tau^+\tau^-$ are seen to be rather small throughout this 
$\tan \beta$ interval while the $gg$ branching fraction remains relatively large, being in the $10^{-3}$ to few $\times 10^{-2}$ range. 
A very important mode in almost all cases (except where it is suppressed by phase space) is $h,H\to AZ$. The reason for this large branching fraction 
is the relative enhancement in the effective $(h,H)AZ$ coupling by a factor of $\sim (M_{h,H}-M_A)/M_Z$ as can be seen from taking the $Z$ Goldstone boson 
limit. In particular, when $A$ is light we see that the mass splitting in the numerator can be quite large ($\sim 500-800$ GeV) relative to $M_Z$.

For our choice of parameters, the decays of the charged Higgs bosons are more straightforwardly understood than those of the corresponding neutral Higgs. 
Clearly, if $M_{H^\pm}$ is in excess of any appropriate pair of fourth generation masses, then these decay modes will dominate, while below this threshold 
decays to $t\bar b$ will be found to dominate. The corresponding partial decay rates to other fermionic final states will be highly suppressed. A 
possibly competing decay mode is $H^\pm \to (h,H,A)W^\pm$ provided phase space is available since it too is somewhat enhanced by the same mechanism discussed above 
in the case of $(h,H)\to AZ$ decay although the mass splittings among the $h,H$ and $H^\pm$ are not always large. 

\begin{figure}[htbp]
\centerline{
\includegraphics[width=8.0cm,angle=90]{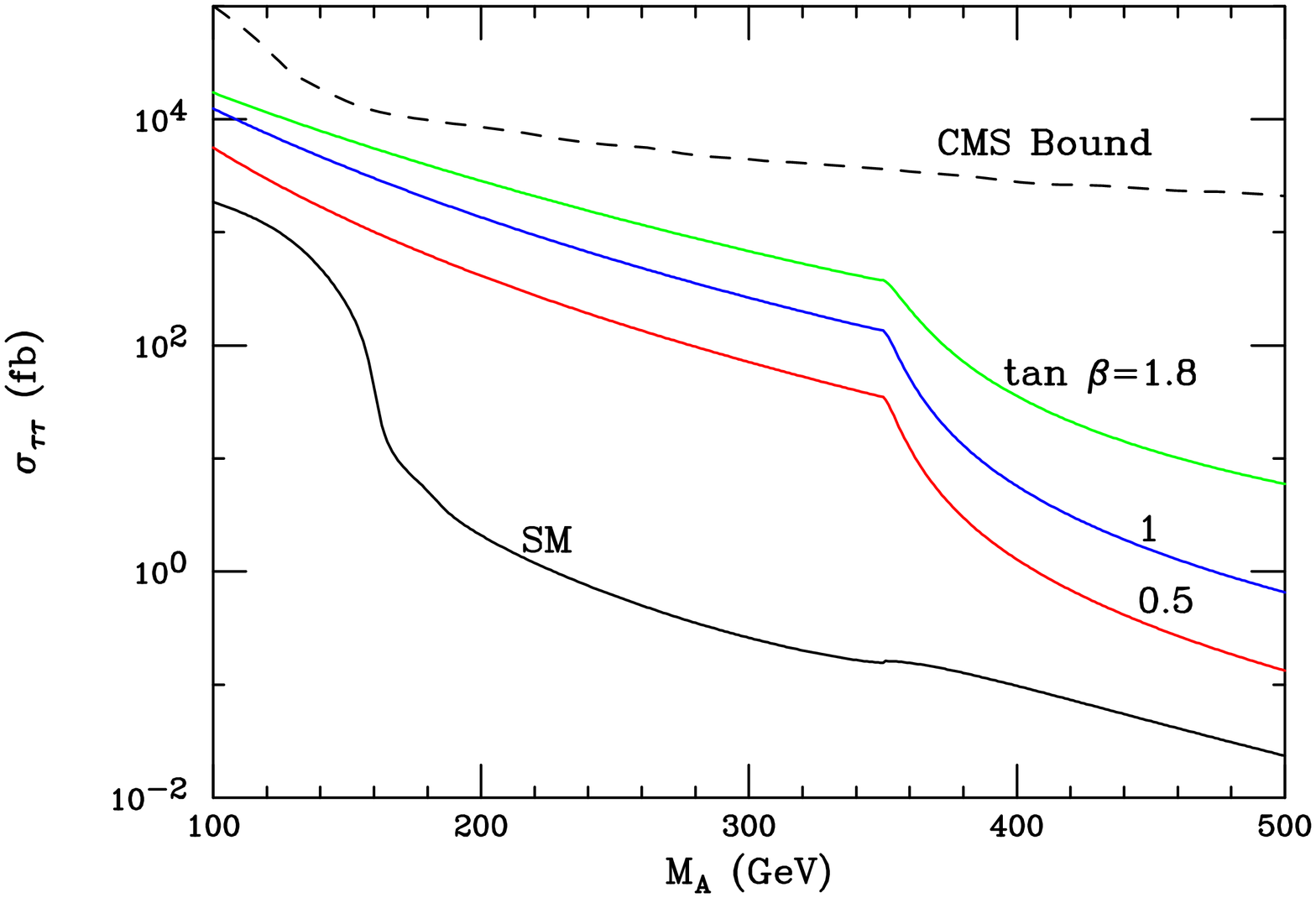}}
\vspace*{0.5cm}
\caption{Cross section times branching fraction for $gg\to A \to \tau \tau$ as a function of $M_A$ for $\tan \beta=$0.5(red), 1(blue) and 1.8(green) 
at the 7 TeV LHC and a comparison to the bound obtained by CMS.}
\label{asigtau}
\end{figure}

Since $A$ is possibly the lightest member of the Higgs spectrum, we first discuss its production signatures at the Tevatron and LHC. Since the $A\to gg$ 
partial width is generally large, $\tan \beta$ is close to unity and the $VVA$ coupling is absent, the $gg\to A$ process is the most important 
one for $A$ production at hadron colliders. These $ggA$ couplings are sufficiently loop-enhanced that one may worry about $gg \to A\to gg$ being seen above 
the dijet background at hadron colliders. Existing searches at the LHC{\cite{Khachatryan:2010jd},\cite{Aad:2011aj}} are only constraining for values of $M_A$ 
beyond our region of interest while those from the Tevatron{\cite{Aaltonen:2008dn},\cite{Abazov:2003tj}} and at lower energies{\cite {UA2}} are found to 
be rather weak. 

For $\tan \beta \gsim 1$, we see from Fig.~\ref{abfs} that the $A\to \tau \tau$ process is a relatively important 
mode but is still subdominant in comparison to both $gg$ and $b\bar b$.  However, the latter two channels are swamped by QCD backgrounds.  
The production cross section for the subprocess $gg\to A\to\tau^+\tau^-$ is shown in Fig.~\ref{asigtau} for $\sqrt s=7$ TeV using the 
CTEQ6.6M parton distribution functions \cite{Nadolsky:2008zw}{\footnote {See the discussion below on how these cross sections are calculated.}}. 
For light $A$ and $\tan \beta \gsim 1$ the resulting cross section is found to be not too far below the (somewhat model-dependent{\cite{Baglio:2011xz}}) 
upper bound recently placed by CMS{\cite{Chatrchyan:2011nx}} as can be seen in Fig.~\ref{asigtau}. However, we note that for smaller values of $\tan \beta$ 
the $gg\to A\to \tau^+\tau^-$ cross section is found to be rather small. 

\begin{figure}[htbp]
\centerline{
\includegraphics[width=6.0cm,angle=90]{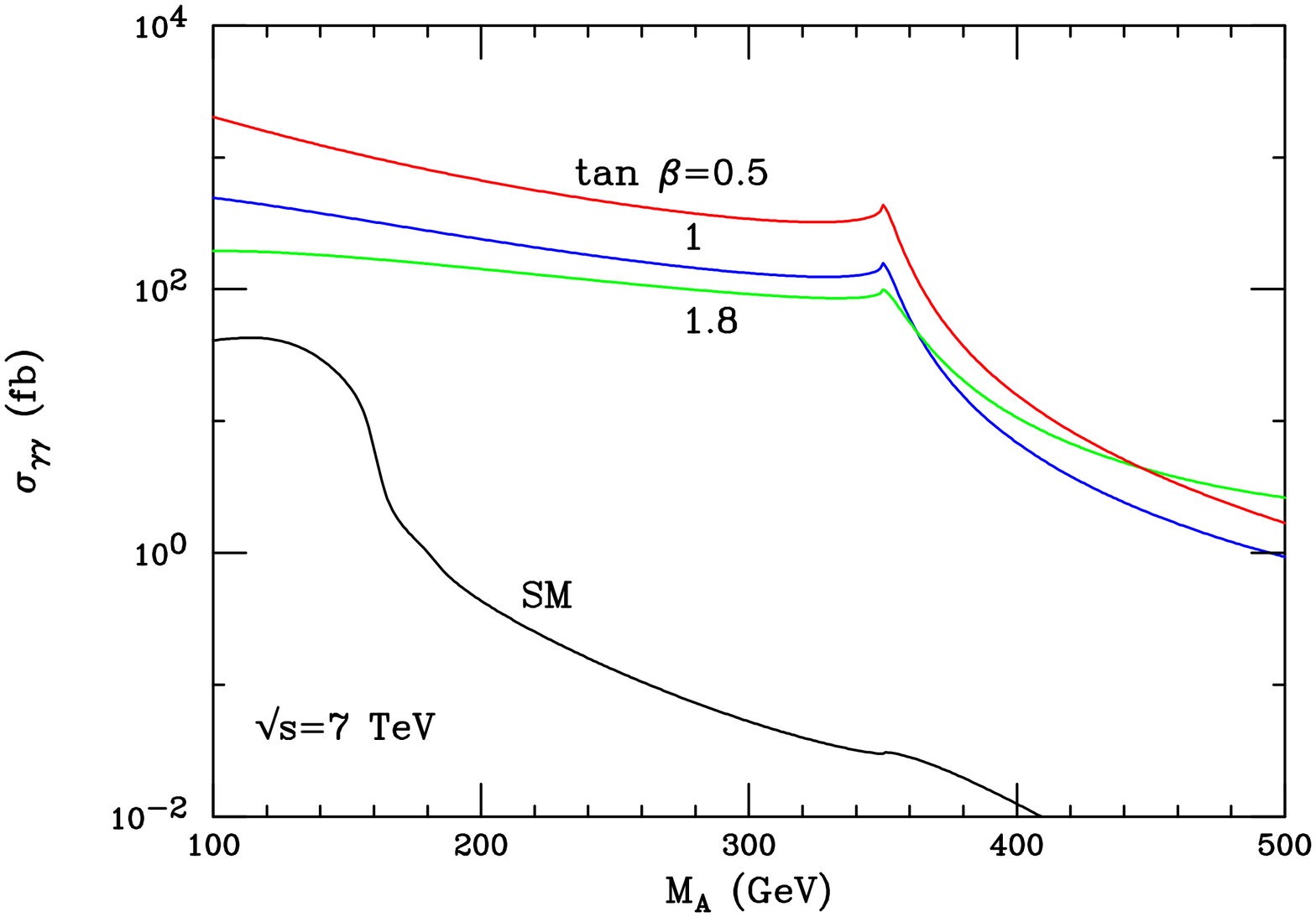}
\hspace*{5mm}
\includegraphics[width=6.0cm,angle=90]{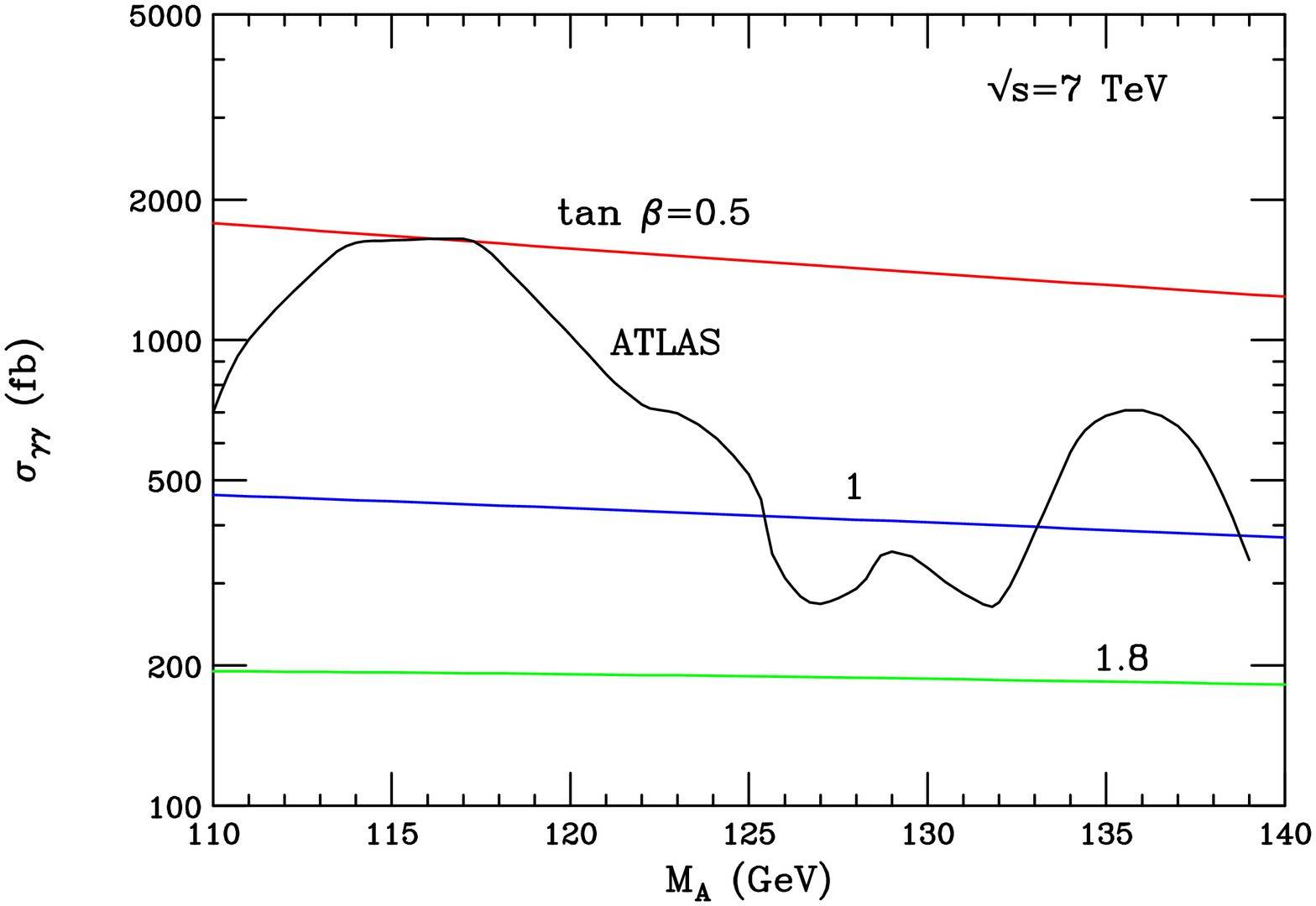}}
\vspace*{0.5cm}
\centerline{
\includegraphics[width=6.0cm,angle=90]{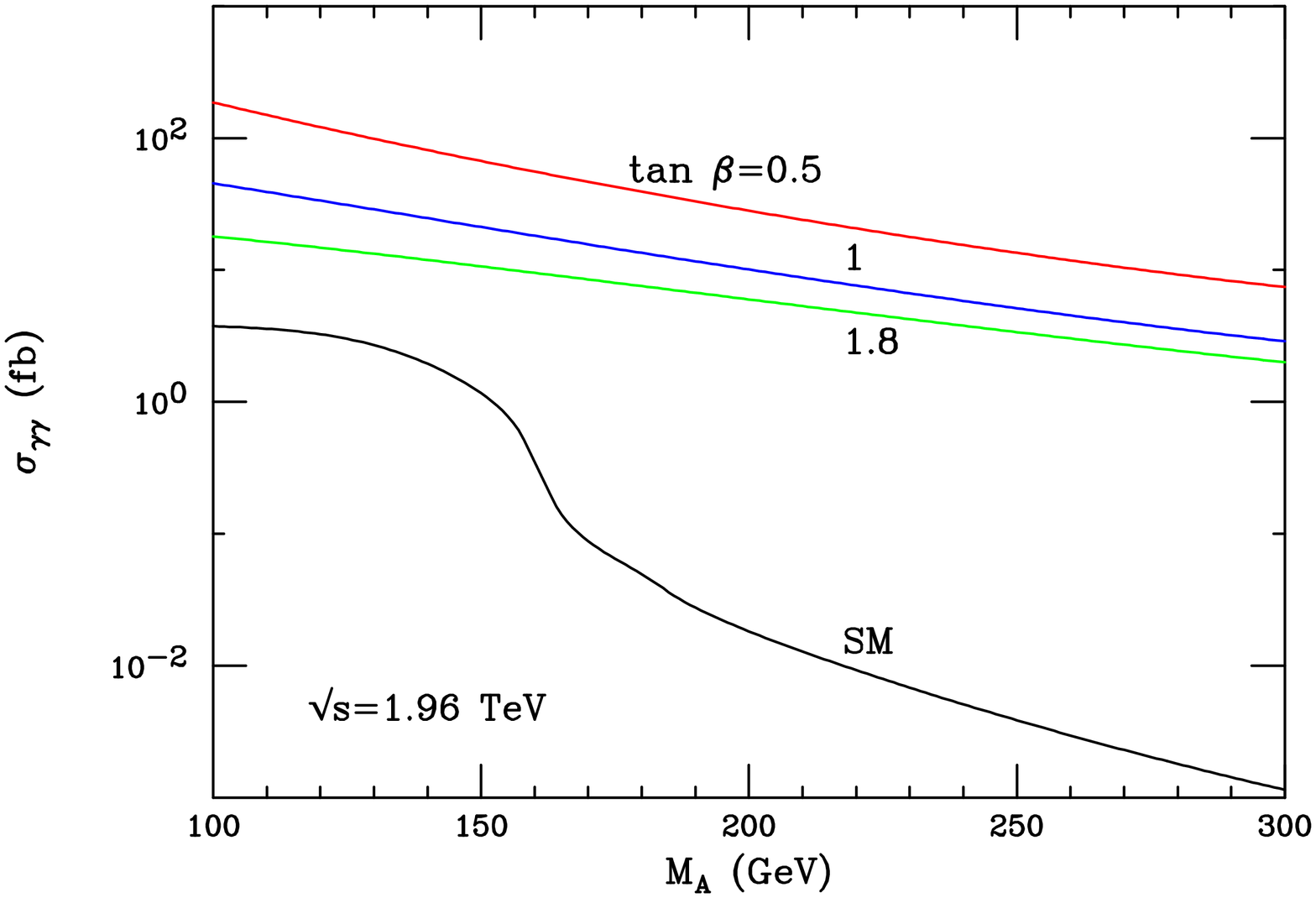}
\hspace*{5mm}
\includegraphics[width=6.0cm,angle=90]{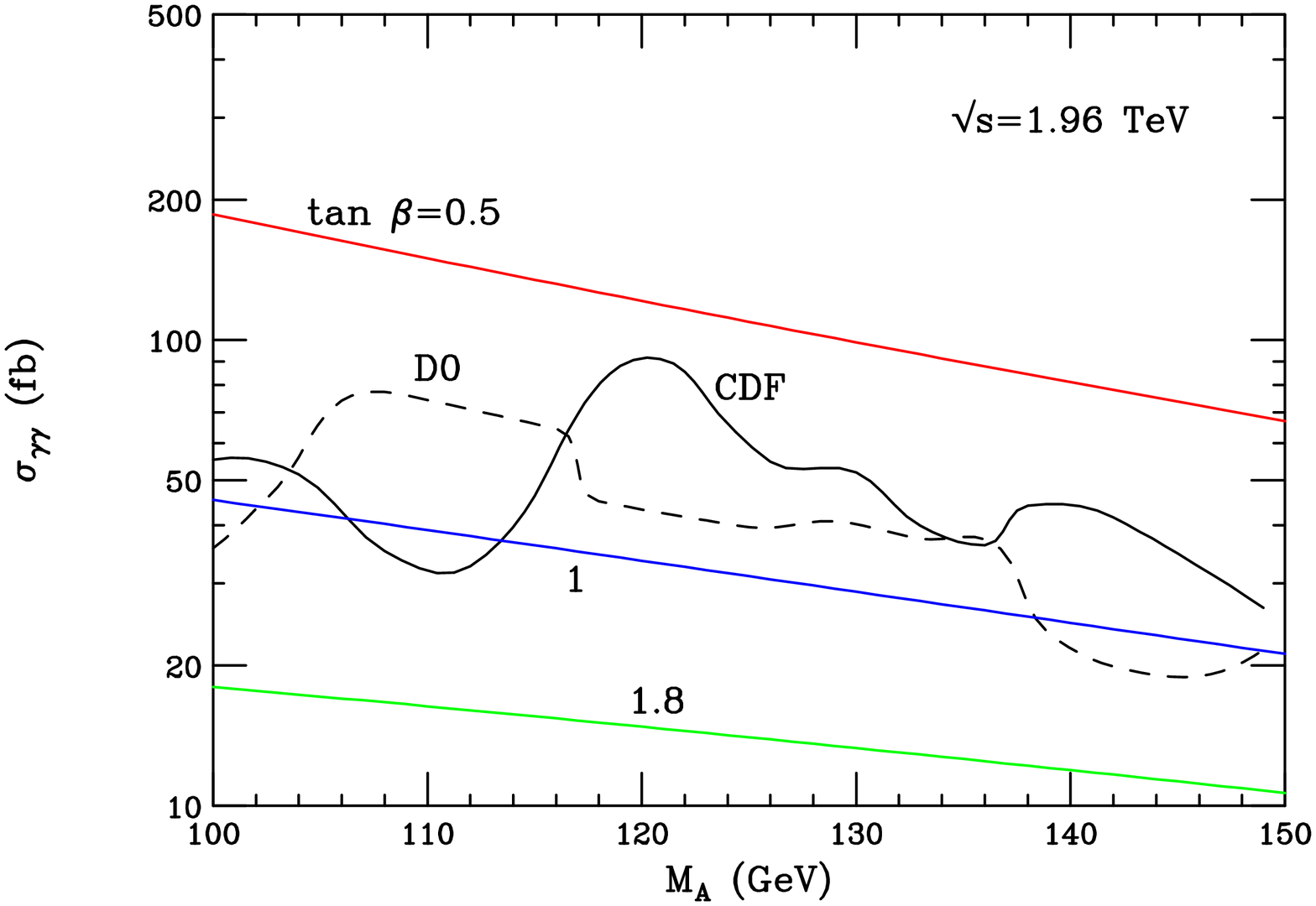}}
\vspace*{0.5cm}
\caption{Cross section times branching fraction for $gg\to A \to \gamma \gamma$ as a function of $M_A$ for $\tan \beta=$0.5(red), 1(blue) and 1.8(green) 
at the 7 TeV LHC(upper left) and Tevatron(lower left). The upper right(lower right) panels explicitly show the limits obtained by CMS at the LHC and by 
CDF and D0 at the Tevatron. The lower solid curve in both left hand panels is the corresponding result for the SM Higgs.}
\label{asig}
\end{figure}

Perhaps the cleanest mode for the observation of a light $A$ boson is in the $\gamma \gamma$ final state; the 7 TeV LHC cross section is shown in 
Fig.~\ref{asig} in comparison to the bound obtained by ATLAS{\cite {ATLASgg}}. We also show in the lower panels the corresponding expectations for the 
Tevatron along with the constraints obtained by both CDF{\cite {CDFgg}} and D0{\cite {{D0gg}}. For this cross section, at either the Tevatron or the LHC, 
we see a significant enhancement for $\tan\beta\lsim 1$. Note that the results shown in this and the previous Figure have 
assumed a constant NNLO K-factor of $\simeq 2$, with the LO cross section renormalized to that for NNLO $A,h$ production for light $A,h$ masses employing the 
results in Ref.{\cite{Dittmaier:2011ti}}. In this approximation our results will give very reasonable overall estimates of the $gg\to A,h,H$ cross sections.   
In this Figure we see that that ratio of cross sections for $gg\to A \to \gamma \gamma$ in comparison to that 
for the corresponding conventional SM Higgs process can be as large as an order of magnitude at lower Higgs masses at hadron colliders.
For example at the LHC, taking $M_A$=115 GeV and $\tan \beta \sim 0.7$ results in a factor $\sim 30$ times larger cross section via the $A$ in this mode than for a 
SM Higgs boson of the same mass. Thus for light $A$ bosons in the range $\sim 100-150$ GeV the $\gamma \gamma$ decay mode may provide the earliest observable 
collider signature.

\begin{table}[htbp]
\centering
\begin{tabular}{|c|c|c|c|} \hline\hline
  & \multicolumn{3}{c|}{$\tan\beta$}\\ \hline
$M_A$ (GeV) & 0.5 & 1.0 & 1.8   \\ \hline
115 & 33.1~(358) & 6.5~(447) & 4.2~(441)\\
300 & 28.9~(375) & 6.3~(467) & 4.4~(454)\\ \hline\hline
\end{tabular}
\caption{Gluon fusion production cross section at the 7 TeV LHC for the lightest
Higgs scalar, $gg\to h+X$, in pb for various values of $M_A$ and
$\tan\beta$.  Fourth generation masses are taken to be $m_{l',\nu'}=300$
GeV, $m_{b'}=350$ GeV, and $m_{t'}=400$ GeV.  The numbers in 
parenthesis indicate the corresponding values of the lightest scalar Higgs
mass in GeV, $m_h$, for these input values of $M_A$ and $\tan\beta$.}
\label{gglighth}
\end{table}

Tables ~\ref{gglighth} and ~\ref{ggheavyh} show the expected $gg$ fusion total cross sections for $h,H$ production obtained by appropriately rescaling the 
NNLO results found in Ref.{\cite{Dittmaier:2011ti}} for some sample values of $M_A$ and $\tan \beta$. While $H$ production in this channel is relatively 
weak due to the larger masses and the reduced effective $ggH$ couplings, $h$ on the other hand is seen to have a substantial cross section with a correspondingly  
respectable branching fraction into both $W^+W^-$ and $ZZ$. For some ranges of these parameters these final states have cross sections that are not very far below 
the present bounds obtained from the 2010 run of the LHC{\cite {atlas,cms}}.

\begin{table}[htpb]
\centering
\begin{tabular}{|c|c|c|c|} \hline\hline
  & \multicolumn{3}{c|}{$\tan\beta$}\\ \hline
$M_A$ (GeV) & 0.5 & 1.0 & 1.8   \\ \hline
115 & 0.14~(849) & 0.58~(543) & 0.78~(645)\\
300 & 0.09~(885) & 0.63~(594) & 0.69~(693)\\ \hline\hline
\end{tabular}
\caption{Gluon fusion production cross section at the 7 TeV LHC for the heaviest
Higgs scalar, $gg\to H+X$, in pb for various values of $M_A$ and
$\tan\beta$.  Fourth generation masses are taken to be $m_{l',\nu'}=300$
GeV, $m_{b'}=350$ GeV, and $m_{t'}=400$ GeV.  The numbers in 
parenthesis indicate the corresponding values of the heaviest scalar Higgs
mass in GeV, $m_H$, for these input values of $M_A$ and $\tan\beta$.}
\label{ggheavyh}
\end{table}

\section{Summary and Conclusions}

In this paper we have examined the properties of the Higgs fields in the 4GMSSM with an eye toward their production signatures at the Tevatron and the LHC.
The couplings and corresponding branching fractions for these various fields were examined in detail. In particular we have noted the strong possibility that 
the CP-odd field $A$ may be the lightest member of the Higgs spectrum as well as the possibility that the region $\tan \beta \lsim 1$ is now physically allowed. 
We further verified that such a light $A$ scenario is consistent with the usual constraints imposed by the electroweak data on the oblique parameters for the 
entire range of perturbatively allowed values of $\tan \beta$. As such, the CP-odd state, $A$, may be the first part of the 4GMSSM Higgs spectrum to be 
discovered at hadron colliders. We find that while $gg\to A$ may soon lead to a potential signal in the $\tau^+\tau^-$ channel at the LHC, $A$ is more likely 
to be first observed in the $\gamma \gamma$ mode due to its highly fourth generation loop-enhanced cross section which can be more than an order of magnitude 
larger than that of the SM Higgs for a mass of $\sim 115-120$ GeV provided that $\tan \beta \lsim 1$. If such a scenario is correct new signals should soon be 
observable at the LHC. 
\bigskip

\noindent{\Large\bf Acknowledgments}\\

AI was supported in part by the Natural Sciences and Engineering Research
Council of Canada.


%
\def\IJMP #1 #2 #3 {Int. J. Mod. Phys. A {\bf#1},\ #2 (#3)}
\def\MPL #1 #2 #3 {Mod. Phys. Lett. A {\bf#1},\ #2 (#3)}
\def\NPB #1 #2 #3 {Nucl. Phys. {\bf#1},\ #2 (#3)}
\def\PLBold #1 #2 #3 {Phys. Lett. {\bf#1},\ #2 (#3)}
\def\PLB #1 #2 #3 {Phys. Lett. B {\bf#1},\ #2 (#3)}
\def\PR #1 #2 #3 {Phys. Rep. {\bf#1},\ #2 (#3)}
\def\PRD #1 #2 #3 {Phys. Rev. D {\bf#1},\ #2 (#3)}
\def\PRL #1 #2 #3 {Phys. Rev. Lett. {\bf#1},\ #2 (#3)}
\def\PTT #1 #2 #3 {Prog. Theor. Phys. {\bf#1},\ #2 (#3)}
\def\RMP #1 #2 #3 {Rev. Mod. Phys. {\bf#1},\ #2 (#3)}
\def\ZPC #1 #2 #3 {Z. Phys. C {\bf#1},\ #2 (#3)}


\begin{thebibliography}{99}


\bibitem{Morrissey:2009tf}
  D.~E.~Morrissey, T.~Plehn, T.~M.~P.~Tait,
  [arXiv:0912.3259 [hep-ph]].

\bibitem{Drees:2004jm}
  M.~Drees, R.~Godbole, P.~Roy,
  Hackensack, USA: World Scientific (2004) 555 p;
  H.~Baer, X.~Tata,
  Cambridge, UK: Univ. Pr. (2006) 537 p.

\bibitem{Hou:2008xd}
  W.~-S.~Hou,
  Chin.\ J.\ Phys.\  {\bf 47}, 134 (2009).
  [arXiv:0803.1234 [hep-ph]];
Y.~Kikukawa, M.~Kohda, J.~Yasuda,
  Prog.\ Theor.\ Phys.\  {\bf 122}, 401-426 (2009).
  [arXiv:0901.1962 [hep-ph]].

\bibitem{Lunghi:2011xy}
  E.~Lunghi, A.~Soni,
 [arXiv:1104.2117 [hep-ph]];
E.~Lunghi, A.~Soni,
  Phys.\ Lett.\  {\bf B697}, 323-328 (2011).
  [arXiv:1010.6069 [hep-ph]];
 W.~-S.~Hou, C.~-Y.~Ma,
  Phys.\ Rev.\  {\bf D82}, 036002 (2010).
  [arXiv:1004.2186 [hep-ph]];
W.~-S.~Hou, Y.~-Y.~Mao, C.~-H.~Shen,
  Phys.\ Rev.\  {\bf D82}, 036005 (2010).
  [arXiv:1003.4361 [hep-ph]];
  M.~Bobrowski, A.~Lenz, J.~Riedl, J.~Rohrwild,
  Phys.\ Rev.\  {\bf D79}, 113006 (2009).
  [arXiv:0902.4883 [hep-ph]];
  S.~K.~Garg, S.~K.~Vempati,
    [arXiv:1103.1011 [hep-ph]].



\bibitem{Litsey:2009rp}
  S.~Litsey, M.~Sher,
  Phys.\ Rev.\  {\bf D80}, 057701 (2009).
  [arXiv:0908.0502 [hep-ph]].

\bibitem{Dawson:2010jx}
  S.~Dawson, P.~Jaiswal,
  Phys.\ Rev.\  {\bf D82}, 073017 (2010).
  [arXiv:1009.1099 [hep-ph]]. For an earlier discussion of the perturbativity constraints and the possibility of Landau poles in 
  the 4GMSSM, see 
  R.~M.~Godbole, S.~K.~Vempati, A.~Wingerter,
  JHEP {\bf 1003}, 023 (2010).
  [arXiv:0911.1882 [hep-ph]].

  



\bibitem{Fok:2008yg}
  R.~Fok, G.~D.~Kribs,
  Phys.\ Rev.\  {\bf D78}, 075023 (2008).
  [arXiv:0803.4207 [hep-ph]].


\bibitem{Kribs:2007nz}
  G.~D.~Kribs, T.~Plehn, M.~Spannowsky, T.~M.~P.~Tait,
  Phys.\ Rev.\  {\bf D76}, 075016 (2007).
  [arXiv:0706.3718 [hep-ph]];
M.~S.~Chanowitz,
  Phys.\ Rev.\  {\bf D}, 035018 (2010).
  [arXiv:1007.0043 [hep-ph]]. See also 
  O.~Eberhardt, A.~Lenz, J.~Rohrwild,
  Phys.\ Rev.\  {\bf D82}, 095006 (2010).
  [arXiv:1005.3505 [hep-ph]]. Fo earlier work in this area, see 
  H.~-J.~He, N.~Polonsky, S.~-f.~Su,
  Phys.\ Rev.\  {\bf D64}, 053004 (2001).
  [hep-ph/0102144].


\bibitem{lepew}
LEP Electroweak Working Group,
http://lepewwg.web.cern.ch/LEPEWWG/


\bibitem{DePree:2009ed}
  E.~De Pree, G.~Marshall, M.~Sher,
  Phys.\ Rev.\  {\bf D80}, 037301 (2009).
  [arXiv:0906.4500 [hep-ph]].

\bibitem{Achard:2001qw}
  P.~Achard {\it et al.}  [L3 Collaboration],
  Phys.\ Lett.\  B {\bf 517}, 75 (2001)
  [arXiv:hep-ex/0107015].

\bibitem{Lister:2011zt}
  A.~Lister [ CDF Collaboration ],
  PoS {\bf ICHEP2010}, 402 (2010).
  [arXiv:1101.5992 [hep-ex]].
  T.~Aaltonen {\it et al.} [ The CDF Collaboration ],
$p\bar{p}$ collisions at $\sqrt{s}=1.96$ TeV,''
  Phys.\ Rev.\ Lett.\  {\bf 106}, 141803 (2011).
  [arXiv:1101.5728 [hep-ex]].

\bibitem{Chatrchyan:2011em}
  S.~Chatrchyan {\it et al.} [ CMS Collaboration ],
  [arXiv:1102.4746 [hep-ex]];
ATLAS Collaboration, ATLAS-CONF-2011-022, March 2011.

\bibitem{CFH}
  M.~S.~Chanowitz, M.~A.~Furman, I.~Hinchliffe,
  Nucl.\ Phys.\  {\bf B153 } (1979)  402  and 
  Phys.\ Lett.\  {\bf B78 } (1978)  285.



\bibitem{ewmoriond}
See talks given by B Jayatilaka (CDF/D0), T. Scanlon (CDF/D0), M. Schumacher (ATLAS), and 
C. Veelken (CMS), at {\it Recontres de Moriond EW 2011}, March 2011, LaThuile, Italy.

\bibitem{hfl}
Heavy Flavor Averaging Group, http://www.slac.stanford.edu/xorg/hfag/;
CKM Fitter, http://ckmfitter.in2p3.fr/.

\bibitem{Aaltonen:2009ke}
  T.~Aaltonen {\it et al.} [ CDF Collaboration ],
  Phys.\ Rev.\ Lett.\  {\bf 103}, 101803 (2009).
  [arXiv:0907.1269 [hep-ex]];
V.~M.~Abazov {\it et al.} [ D0 Collaboration ],
  Phys.\ Lett.\  {\bf B682}, 278-286 (2009).
  [arXiv:0908.1811 [hep-ex]].




\bibitem{Sher:1988mj}
  M.~Sher,
  Phys.\ Rept.\  {\bf 179}, 273-418 (1989).

\bibitem{Barger:1991ed}
  V.~D.~Barger, M.~S.~Berger, A.~L.~Stange, R.~J.~N.~Phillips,
  Phys.\ Rev.\  {\bf D45}, 4128-4147 (1992).

\bibitem{Peskin:1991sw}
  M.~E.~Peskin, T.~Takeuchi,
  Phys.\ Rev.\  {\bf D46}, 381-409 (1992).





\bibitem{He:2001tp}
  H.~J.~He, N.~Polonsky and S.~f.~Su,
  Phys.\ Rev.\  D {\bf 64}, 053004 (2001)
  [arXiv:hep-ph/0102144].

\bibitem{pdg}
K. Nakamura et al. (Particle Data Group), J. Phys. G 37, 075021 (2010);
GFitter Working Group, http://gfitter.desy.de/.



\bibitem{Khachatryan:2010jd}
  V.~Khachatryan {\it et al.} [ CMS Collaboration ],
  Phys.\ Rev.\ Lett.\  {\bf 105}, 211801 (2010).
  [arXiv:1010.0203 [hep-ex]].

\bibitem{Aad:2011aj}
  G.~Aad {\it et al.} [ ATLAS Collaboration ],
    [arXiv:1103.3864 [hep-ex]].

\bibitem{Aaltonen:2008dn}
  T.~Aaltonen {\it et al.} [ CDF Collaboration ],
  Phys.\ Rev.\  {\bf D79}, 112002 (2009).
  [arXiv:0812.4036 [hep-ex]].

\bibitem{Abazov:2003tj}
  V.~M.~Abazov {\it et al.} [ D0 Collaboration ],
  Phys.\ Rev.\  {\bf D69}, 111101 (2004).
  [hep-ex/0308033].


\bibitem{UA2}
  J.~Alitti {\it et al.} [ UA2 Collaboration ],
  Z.\ Phys.\  {\bf C49 } (1991)  17-28 and 
  Nucl.\ Phys.\  {\bf B400 } (1993)  3-24.
  

\bibitem{Nadolsky:2008zw}
  P.~M.~Nadolsky, H.~-L.~Lai, Q.~-H.~Cao, J.~Huston, J.~Pumplin, D.~Stump, W.~-K.~Tung, C.~-P.~Yuan,
  Phys.\ Rev.\  {\bf D78 } (2008)  013004.
  [arXiv:0802.0007 [hep-ph]].


\bibitem{Baglio:2011xz}
  J.~Baglio, A.~Djouadi,
    [arXiv:1103.6247 [hep-ph]].


\bibitem{Chatrchyan:2011nx}
  S.~Chatrchyan {\it et al.} [ CMS Collaboration ],
    [arXiv:1104.1619 [hep-ex]].

\bibitem{ATLASgg}
ATLAS Collaboration, ATLAS-CONF-2011-025.

\bibitem{CDFgg}
CDF Collaboration, CDF/PUB/EXOTIC/PUBLIC/10485

\bibitem{D0gg}
D0 Collaboration, D0 Note 6177-CONF

\bibitem{Dittmaier:2011ti}
  S.~Dittmaier {\it et al.} [ LHC Higgs Cross Section Working Group Collaboration ],
    [arXiv:1101.0593 [hep-ph]].


\bibitem{atlas}
See ATLAS notes ATLAS-CONF-2011-048 for Higgs searches in the $ZZ$ channel and ATLAS-CONF-2011-052 for those in $W^+W^-$ channel.

\bibitem{cms}
  S.~Chatrchyan {\it et al.} [ CMS Collaboration ],
  Phys.\ Lett.\  {\bf B699}, 25-47 (2011).
  [arXiv:1102.5429 [hep-ex]].




\end{thebibliography}
\end{document}